\documentclass[a4paper,10pt]{article}

\usepackage{fullpage} % Package to use full page
\usepackage{parskip} % Package to tweak paragraph skipping
\usepackage{tikz} % Package for drawing
\usepackage{amsmath}
\usepackage{hyperref}
\usepackage{graphicx}
\usepackage[font=footnotesize]{caption}
\usepackage{array}
\usepackage{multicol}
\usepackage{amssymb}
\usepackage[ruled,vlined,linesnumbered]{algorithm2e}
\usepackage{mathrsfs}
\usepackage{bm}
\SetAlFnt{\footnotesize}
\usepackage{adjustbox}
\graphicspath{ {images/} }
\usepackage{color}
\usepackage{hyperref}
\usepackage{booktabs}
\usepackage{multirow}
\usepackage[sort]{natbib}
\usepackage{numprint}
\usepackage{tikz}
\usetikzlibrary{bayesnet}
\usetikzlibrary{arrows}
\usepackage{subcaption}
\usepackage{subfiles}
\usepackage{mathtools}
\usepackage{breqn}
\usepackage{lscape}
\usepackage[prependcaption,colorinlistoftodos]{todonotes}
\usepackage{authblk}
\usepackage{booktabs}

\newcommand{\x}{{\bm{x}}}
\newcommand{\z}{{\bm{z}}}
\newcommand\xo{\bm{x}_{\scriptstyle\scriptscriptstyle\mathcal{O}}}
\newcommand\Xo{\bm{X}_{\scriptstyle\scriptscriptstyle\mathcal{O}}}
\newcommand\xm{\bm{x}_{\scriptstyle\scriptscriptstyle\mathcal{M}}}

\renewcommand\abstractname{\textsc{ABSTRACT}}

\newcommand{\review}[1]{\textcolor{black}{#1}}

\begin{document}

\title{Multimodal Generative Models for \\ Bankruptcy Prediction Using Textual Data}

\author[a]{\normalsize{Rogelio A. Mancisidor$^{\ast}$}}
\author[b]{\normalsize{Kjersti Aas}}
\affil[a]{\normalsize{Department of Data Science \& Analytics BI Norwegian Business School. Nydalsveien 37, 0484 Oslo Norway.} E-mail address: \href{mailto:rogelio.a.mancisidor@bi.no}{rogelio.a.mancisidor@bi.no}}
\affil[b]{\normalsize{Norwegian Computing Center, P.O. Box 114 Blindern, Oslo Norway.} E-mail address: \href{mailto:kjersti@nr.no}{kjersti@nr.no}}
%\affil[a]{\href{mailto:rogelio.a.mancisidor@bi.no}{rogelio.a.mancisidor@bi.no}} 
%\affil[b]{\href{mailto:kjersti@nr.no}{kjersti@nr.no}}
\affil[*]{Corresponding author}

\maketitle

\renewenvironment{abstract}
{\begin{quote}
\noindent \rule{\linewidth}{.5pt}\par{\bfseries \abstractname.}}
{\noindent 
\rule{\linewidth}{.5pt}
\end{quote}
}

\begin{abstract}
Textual data from financial filings, e.g., the Management's Discussion \& Analysis (MDA) section in Form 10-K, has been used to improve the prediction accuracy of bankruptcy models. In practice, however, we cannot obtain the MDA section for all public companies, \review{which limits the use of MDA data in traditional bankruptcy models, as they need complete data to make predictions.} The two main reasons for the lack of MDA are: (i) not all companies are obliged to submit the MDA and (ii) technical problems arise when crawling and scrapping the MDA section. \review{To solve this limitation, this research introduces the Conditional Multimodal Discriminative (CMMD) model that learns multimodal representations that embed information from accounting, market, and textual data modalities.} The CMMD model needs a sample with all data modalities for model training. At test time, the CMMD model only needs access to accounting and market modalities to generate multimodal representations, which are further used to make bankruptcy predictions \review{and to generate words from the missing MDA modality}. \review{With this novel methodology, it is realistic to use textual data in bankruptcy prediction models}, since accounting and market data are available for all companies, unlike textual data. The empirical results of this research show that \review{if financial regulators, or investors, were to use traditional models using MDA data, they would only be able to make predictions for 60\% of the companies.} Furthermore, the classification performance of our proposed methodology is superior to that of a large number of traditional classifier models, \review{taking into account all the companies in our sample}.

\end{abstract}

\section{Introduction}\label{sec_intro}
Corporate bankruptcy prediction plays an important role for both regulators and analysts in the financial industry \citep{ding_class_2012}. Therefore, there is a vast body of literature on bankruptcy models (see Table \ref{tbl_overview}), which mostly use panel data containing accounting and market information to predict whether a company will fall into financial distress. Most recently, textual data from financial filings, such as the Management's Discussion \& Analysis (MDA) section of Form 10-K, has been used to improve the prediction accuracy of bankruptcy models, e.g.  \cite{cecchini2010making,mayew2015md,mai_deep_2019,kim2021corporate} and \cite{chen2023bankruptcy}, as it provides a forward-looking view of the company's performance. The information contained in the MDA section has also been used in other research domains, e.g., forecasting corporate investment \citep{cho2021firms,berns2022changes} or financial events \citep{cecchini2010making}, spillover effects \citep{durnev2020spillover}, or in tax avoidance \citep{zhang2022defend}, just to name a few. In all cases the MDA section has proven to contain valuable information.

Traditional methods for predicting corporate bankruptcy require the same data to be used for training the model and for making new bankruptcy predictions. Unfortunately, for models using the MDA section this is not always possible. While it is true that the financial regulator requires a wealth of information in the company's annual reports on Form 10-K, not all companies are obliged to fulfill this requirement\footnote{Reporting companies can be a company which is listed on a Public Exchange or not listed on an exchange but traded publicly. If a company's total asset amounts to more than 10 million USD and it has a class of equity securities that is held of record either by 2,000 or more persons or by 500 or more non-accredited investors then it is obligated to file a registration statement under Section 12 of the Securities Exchange Act of 1934. Otherwise, companies are not obliged to file annual or quarterly reports.}. Additionally, obtaining the MDA section for statistical modeling involves technical procedures (Web crawling and scraping) that do not guarantee a successful extraction. As a consequence, using textual data in bankruptcy models to develop relatively more accurate models is not feasible in practice,  since financial regulators or investors are not able to make bankruptcy predictions for all companies.

To solve the above limitation, this research introduces a novel methodology for bankruptcy prediction, which is based on multimodal learning and uses the Conditional MultiModal Discriminative (CMMD) model introduced in \cite{mancisidor2021discriminative}. Under the CMMD framework, accounting ($\x_1$), market ($\x_2$), and textual data ($\x_3$) are considered data modalities that provide different information about the financial condition of a given company. Further, the CMMD framework assumes that accounting and market modalities are always observed, i.e., $\xo = (\x_1,\x_2)$. On the other hand, textual data $\xm = \x_3$ and class labels $y$ are available for model training but missing at test time\footnote{The subscripts in $\xo$ and $\xm$, therefore, indicate whether data modalities are \textbf{o}bserved or \textbf{m}issing at test time.}. After optimization of the objective function, the CMMD model embeds information from all data modalities ($\xo$ and $\xm$) into a multimodal data representation ($\z$). The new  representation $\z$ of the data modalities is believed to be capable of capturing the posterior distribution of the explanatory factors of the data modalities $\xo$ and $\xm$, and is therefore useful as input to a classifier model \citep{bengio2013representation}. Hence, the CMMD model can predict corporate bankruptcy using multimodal representations $\z$, instead of the accounting, market, and textual data themselves.

The CMMD model requires all data modalities and class labels only during model training. At test time, CMMD generates multimodal representations $\z$ for a new company simply by using $\xo$, i.e., accounting and market data. This is possible by minimizing a divergence measure between a prior distribution $p(\z|\xo)$ (conditioned on the always observed data modalities) and a posterior distribution $q(\z|\xo,\xm,y)$ (conditioned on all data modalities and class labels). Such minimization has the effect of bringing the prior close to the posterior \citep{suzuki2016joint}, and it also minimizes the expected information required to convert a sample from the prior into a sample from the posterior distribution \citep{doersch2016tutorial}. Using multimodal representations $\z \sim p(\z|\xo)$ for bankruptcy prediction, and not the data modalities $\xo$ and $\xm$ themselves, has therefore one main advantage. That is, the CMMD model avoids the limitation of previous bankruptcy models since CMMD does not require the information contained in $\xm$ to be available for making new bankruptcy predictions. 
%\review{Therefore, the focus of this research is not on assessing the predictive power of different methodologies to transform the MDA section into a numerical format that can be used in classifier models, e.g. dictionary-based approach, word2vec, sentiment analysis, or term-frequency, matrices.} 

Using data from companies listed on the AMEX, NYSE and NASDAQ stock exchanges, the empirical results of this research show that for relatively large training data sets, our proposed methodology achieves higher classification performance compared to traditional classifiers, which use accounting and market modalities to make bankruptcy predictions. \review{If financial regulators, or investors, were to use traditional models using MDA data, they would only be able to make predictions for 60\% of the companies according to our data sample, as 40\% of the companies in our data lack MDA for one of the reasons explained above.} On the contrary, our proposed methodology for bankruptcy prediction, which only use accounting and market modalities to generate multimodal representations, may be used for all companies.
To summarize, our main contributions are:
\begin{itemize}
    \item We resolve the limitation of previous bankruptcy models that require MDA data to make predictions, as they can only make predictions for a proportion of firms that is significantly smaller than the number it would need to be in reality. This makes the use of MDA data realistic and possible.
    
    \item We introduce for the first time, to the best of our knowledge, the concept of multimodal learning for corporate bankruptcy models.
    %\item We propose an index that is based on multimodal representations and able to capture the uncertainty of the financial situation of companies.
\end{itemize}

\section{Related Work}\label{sec_related}
Since the seminal work of \cite{beaver1966financial} and \cite{altman_financial_1968} on corporate bankruptcy prediction, research in this field has grown rapidly over the past 50 years. Therefore, this section focuses on the application of neural networks and textual data for bankruptcy prediction. For an exhaustive review on other bankruptcy prediction models, the reader is referred to \cite{kumar2007bankruptcy,demyanyk2010financial} and \cite{alaka2018systematic}.

The use of neural networks (NNs) in bankruptcy prediction dates back to 1990 with the research by \cite{bell1990neural} and \cite{odom1990neural}. The authors compare the relative performance of NNs over logistic regression (LR) and multivariate discriminant analysis (DA), respectively. In both cases, the architecture of the NN is simple. \cite{bell1990neural} use one hidden layer with 6 neurons, while \cite{odom1990neural} use 5 neurons in the hidden layer. In both cases, NNs are optimized with the backpropagation algorithm \citep{rumelhart1985learning}. The results in these pioneering studies are contradictory; NNs show no significant improvement over LR \citep{bell1990neural}, but outperform DA in predicting bankruptcy \citep{odom1990neural}. 

\begin{table}[t!]
\centering 
\caption{Research on bankruptcy prediction with neural networks. The column "Benchmark models" includes the following abbrevations: logistic regression (LR), probit regression (PR), discriminant analysis (DA), linear classifier (LC), k-nearest neighbors (kNN), decision trees (DT), random forest (RF), probabilistic neural networks (PNN), self organizing maps (SOM), genetic algorithm (GA), generalized additive models (GAM), ensemble models (EM), fuzzy neural networks (FNN), stacked generalization model (SGM), support vector machine (SVM), leave-one-out incremental extreme learning machine (LOO-IELM), XGBoost (XGB), and deep learning (DL). }
\def\arraystretch{1.1}
\begin{adjustbox}{width=\textwidth}
\begin{tabular}{|l|l|c|c|c|c|}
\hline
 (Year) \  Author & Benchmark Models & No. obs. & No. bankruptcies & Period & No. years \\ %Improvement \\
\hline
(1990) Bell et al. \citep{bell1990neural}  & LR   & 2067 & 233 & 1985-1986 & 2  \\ % & On a par \\
(1990) Odom and Sharda \citep{odom1990neural} &   DA   &   129&65&1975-1982&8\\ % & On a par \\&Y\\
(1992) Tam and Kiang \citep{tam1992managerial} &   DA, LC, LR, kNN, and DT &162&81&1985-1987&3\\ % & On a par \\&Y\\
(1992) Salchenberger et al. \citep{salchenberger1992neural} & LR & 316, 404 & 158, 75 &1986-1987&2\\ % & On a par \\&Y\\
(1993) Chung and Tam \citep{chung1993comparative} & DA, LR, DT    &162&81&1985-1987&3\\ % & On a par \\&Y\\
(1993) Coats and Fant \citep{coats1993recognizing}  & DA    &   282&94&1970-1989&20\\ % & On a par \\&Y\\
(1993) Fletcher and Goss \citep{fletcher1993forecasting} &  LR  &   36&18&1971-1979&9\\ % & On a par \\&Y\\
(1994) Wilson and Sharda \citep{wilson1994bankruptcy}  &   DA  &129&65&1975-1982&8\\ % & On a par \\&Y\\
(1994) Fanning and Cogger \citep{fanning1994comparative}  & LR &380&190&1942-1965&24\\ % & On a par \\&Y\\
(1995) Boritz et al. \citep{boritz1995predicting} &  DA, LR, PR  &6324&171&1971-1984&14\\ % & On a par \\&On a pair\\
(1997) Etheridge and Sriram \citep{etheridge1997comparison} &   DA, LR  &1139&148&1986-1988&3\\ % & On a par \\&When error costs are concidered\\
(1997) Barniv et al. \citep{barniv1997predicting} &  DA, LR  &237&69&1980-1991&12\\ % & On a par \\&In a subsample of continously traded firms\\
(1997) Jain and Nag \citep{jain1997performance} & LR &  431&327&1976-1988&13\\ % & On a par \\&Y\\
(1999) Yang et al. \citep{yang1999probabilistic} & DA, PNN & 122&33&1984-1989&6\\ % & On a par \\&Y\\
(1999) Zhang et al. \citep{zhang1999artificial}  & LR &220&110&1980-1991&12\\ % & On a par \\&Y\\
(2001) Atiya \cite{atiya2001bankruptcy}  &   -   &   1160&444&-&-\\ % & On a par \\&Y\\
(2005) Lee et al. \citep{lee2005comparison}  &   SOM, DA, LR &   168&84&1995-1998&4\\ % & On a par \\&Y\\
(2005) Pendharkar \citep{pendharkar2005threshold}  &   DA, GA, DT   &200&100&1987-1995&9\\ % & On a par \\&Y\\
(2006) Neves and Vieira \citep{neves2006improving} &   DA    &   2800&583&1998-2000&3\\ % & On a par \\&Y\\
(2006) Ravikumar and Ravi \citep{ravikumar2006bankruptcy} & -&129&65&-&-\\ % & On a par \\&-\\
(2006) Berg \citep{berg2007bankruptcy} &   GAM, DA, &-&-&1996-2000&5\\ % & On a par \\&N\\
(2008) Tsai and Wu \citep{tsai2008using}  & EM & 690, 1000, 360&383, 300, 383&-&-\\ % & On a par \\&-\\
(2008) Alfaro et al. \citep{alfaro2008bankruptcy} & EM, DA, DT    &1180&590&2000-2003&4\\ % & On a par \\&-\\
(2008) Ravi and Pramodh \citep{ravi2008threshold}  &  - &  66, 40&-&-&-\\ % & On a par \\&-\\
(2008) Huang \citep{huang2008genetic}  &   FNN &   400&80&2002-2005&4\\ % & On a par \\&-\\
(2009) Chauhan et al. \citep{chauhan2009differential} & WNN, DEWNN, TAWNN & 40, 66, 129&22, 37, 65&-&-\\ % & On a par \\&-\\
(2010) Kim and Kang \citep{kim2010ensemble} &   EM  &1458&729&2002-2005&4\\ % & On a par \\&-\\
(2010) Cecchini et al. \citep{cecchini2010making} &-&156&78&1994-1999&6\\
(2012) Jeong et al. \citep{jeong2012tuning} &   GAM   &   2542&1271&2001-2004&4\\ % & On a par \\&-\\
(2013) Tsai and Hsu \citep{tsai2013meta} &   SGM   &-&-&-&-\\ % & On a par \\&-\\
(2014) Tsai et al. \citep{tsai2014comparative} &   EM, DT, SVM &   690, 1000, 690&383, 300, 383&-&-\\ % & On a par \\&-\\
(2014) Yu et al. \citep{yu2014bankruptcy} &   LOO-IELM, EM &   1020&520&2002-2003&2\\ % & On a par \\&-\\
(2015) Iturruaga and Sanz  \citep{iturriaga2015bankruptcy} & DA, LR, SVM, RF & 772&386&2002-2012&11\\ % & On a par \\&Y\\
(2015) Mayew et al. \citep{mayew2015md} &-&45725&460&1995-2012&18\\
(2016) Zieba et al. \citep{zikeba2016ensemble} &-&10700&700&2007-2013&7\\ % & On a par \\&-\\
(2019) Mai et al. \citep{mai_deep_2019} & DL, LR, SVM, RF   & 11827&477&1994-2014&21\\ % & On a par \\&-\\
(2023) Chen et al. \cite{chen2023bankruptcy} & LR, RF, SVM, XGB &   41439   & 932   &1994-2018 & 25\\
%Ours                                         & kNN, NB, LR, SVM, RF, MLP &                     
\hline
\end{tabular}
\end{adjustbox}
\label{tbl_overview}
\end{table}

There are several papers comparing the performance of NNs and traditional models. For example,  \cite{chung1993comparative, coats1993recognizing, wilson1994bankruptcy,zhang1999artificial} and \cite{lee2005comparison} use a NN with a single hidden layer and compare its performance with that of decision trees (DT), DA, LR, and self-organizing maps (SOMs). \cite{yang1999probabilistic} compare NNs with probabilistic NN and their results show that both models are equally good for bankruptcy prediction. In the early research during the 1990s, some of the concerns with NNs were: i) the lack of a formal method to choose the NN architecture, ii) computationally demanding training, and iii) the lack of model interpretation \citep{tam1992managerial,salchenberger1992neural,barniv1997predicting}. To provide solutions to these problems, \cite{fanning1994comparative} propose an adaptive NN to choose the network architecture and both \cite{fletcher1993forecasting} and \cite{yu2014bankruptcy} propose methods to select the number of neurons in the hidden layer; the former use a grid search approach, while the latter use a method called leave one out incremental extreme learning machine. Furthermore, \cite{huang2008genetic} use the genetic algorithm to add decision rules for model interpretability and in \cite{jeong2012tuning} to select the number of hidden units and weight decay in NNs. 

Given that real bankruptcy data are highly imbalanced, \cite{tam1992managerial} and \cite{fanning1994comparative} use prior probabilities and \cite{etheridge1997comparison} use relative costs to take into account misclassification costs. Both \cite{boritz1995predicting} and \cite{jain1997performance} vary the number of non-bankruptcy firms to assess the impact of class imbalance, and \cite{zikeba2016ensemble} generate synthetics data to improve the classification results.

SOMs and competitive-learning are coupled with NNs in \cite{etheridge1997comparison} and \cite{iturriaga2015bankruptcy}. \cite{etheridge1997comparison} consider different forecast horizons and their results show an increase in the relative classification performance of NNs as the forecast horizon increases. \cite{etheridge1997comparison} argue that models that are able to predict bankruptcy 2 or 3 years ahead can be used as early warning support systems by financial authorities. \cite{neves2006improving} apply a supervised learning vector quantization to the last hidden layer of the NN, aiming to correct the errors produced by the NN.  

Ensemble models are introduced in \cite{ravikumar2006bankruptcy,tsai2008using,alfaro2008bankruptcy,kim2010ensemble,tsai2014comparative} and \cite{zikeba2016ensemble}. In \cite{tsai2008using} different NNs are ensembled, while \cite{ravikumar2006bankruptcy} test the performance of 7 different ensemble models, and \cite{alfaro2008bankruptcy} use adaboost as the learning method. Both \cite{kim2010ensemble} and \cite{tsai2014comparative} compare bagging and boosting learning methods, but the latter varies the number of models from 10 to 100. Boosting is also used in \cite{zikeba2016ensemble}, in the form of extreme gradient boosting. Only adaboost, bagging, and boosting ensemble methods outperform the benchmark models under study. 

Some research has focused on the data used to predict bankruptcy, rather than the model itself. In both \cite{atiya2001bankruptcy} and \cite{jeong2012tuning} the focus is on selecting the input features. Specifically, \cite{atiya2001bankruptcy} tests the predictive power of stock information, and \cite{jeong2012tuning} introduce a generalized additive model (GAM) that selects the best input features. In \cite{barniv1997predicting} the bankruptcy definition is a three-state random variable, i.e., acquisition, emerging as independent entities, or liquidated. Finally, \cite{berg2007bankruptcy} focuses on non-linearities in the input data, introducing a generalized additive model that uses a sum of smooth functions to model potential non-linear shapes of covariate effects. 

All \cite{pendharkar2005threshold,ravi2008threshold,chauhan2009differential} and \cite{tsai2013meta} introduce novel ideas for bankruptcy prediction. \cite{pendharkar2005threshold} trains a NN and the classification threshold end-to-end, i.e., the threshold is trained simultaneously with the weights of the NN, so that accuracy is maximized. To reduce the number of weights in NNs, \cite{ravi2008threshold} replace the hidden layer in a regular NN for principal components and train such an architecture using the threshold accepting algorithm. Another interesting NN architecture is the wavelet neural network (WNN) that is presented in \cite{chauhan2009differential}. Those authors use the evolution algorithm to train a WNN, which is relatively more robust to variations in the hyperparameters, so parameter tuning is relatively easy. Finally, \cite{tsai2013meta} present a meta-learning approach for bankruptcy prediction in which two-level classifiers are employed. The first level, composed by different classifiers, filters out irrelevant data. The second level, composed by a single classifier making the final predictions, is trained by the data from the first level. 

The first research that uses the MDA section for bankruptcy prediction is presented by \cite{cecchini2010making}. The authors create a dictionary of key terms associated with the bankruptcy event using computational linguistic theory, while \cite{mayew2015md} search for sentences explicitly referring to the term ``going concern" to create a binary variable and look into the linguistic tone of the MDA section. Language models are used in \cite{kim2021corporate}, where the BERT model \citep{devlin2018bert} is used to find the sentiment of the MDA section. Then sentiment is further used to predict bankruptcy. In a different approach, \cite{mai_deep_2019} convert the MDA into numerical vectors using the skip-gram model \cite{mikolov2013distributed} and the Term Frequency-Inverse Document Frequency matrix (TF-IDF). Then, the vectors obtained with the skip-gram model are the input for an average embedding model and convolutional NN and the TF-IDF vectors are used in the benchmark models. In all cited papers, the authors conclude that the MDA section contains information that is useful to predict bankruptcy. Based on this previous literature, we can see that there are different methods to transform the content of the MDA into a numerical vector that can be used in classifier models. The focus of our research is on solving the limitation of classical methodologies that only can make predictions for a small proportion of companies, and not on assessing which of the previous methods to transform the MDA work best.

\review{Recently, \cite{chen2023bankruptcy} use readability and tone variables\footnote{The methodology for the readability and tone variables is originally introduced in \cite{seebeck2023power}.} derived from the firm's Form 10-K, in addition to financial variables, to improve the performance of corporate bankruptcy predictions. The readability and tone variables are available in the SEC Analytics Suite Database, where 36 variables can be obtained. Of those 36 variables, \cite{chen2023bankruptcy} selects only 11 variables, which are uncorrelated and with the highest predictive power. Then, to forecast 1-year bankruptcies in 2015, for example, they use trained models with data from 1994 to 2014. Likewise, 1-year bankruptcy predictions in 2016 are done with models trained with data from 1994 to 2015, i.e. their methodology is based on the assumption that it is possible to retrain models every year. Therefore, neither the textual data nor the methodology are similar to those of this research. Our proposed methodology, on the other hand, uses the actual MDA text to test the generative capabilities of the CMMD model, and the bankruptcy predictions are not done on the premise that we need to retrain the models every year, i.e. if we train the CMMD model with data from 1994 to 2014, this trained model is used to predict bankruptcies in both 2015 and 2016, for example. Therefore, our experimental setup considers a relatively challenging scenario. 
}
%KJERSTI: Vet ikke helt om jeg skj\o nte hva du mener med "readability and tone variables"? Lager de seg noen variable fra teksten og i saafall hvor mange?  Jeg skjonte heller ikke helt hva du mener med at de bruker en "rolling approach". Bruker ikke du ogsaa en rolling approach? Klarer du aa skrive forskjellen mellom hva de og du gjoer litt klarere?  

\section{Methods}\label{sec_methods}
This section discusses the different methodologies used in our multimodal approach for predicting bankruptcy. For a comprehensive review of the methods presented in this section, the reader is referred to \cite{baltruvsaitis2018multimodal,blei2017variational,kingma2013auto} and \cite{mancisidor2021discriminative}.

\subsection{Multimodal learning}
\begin{figure}[!t]
    \centering
    \includegraphics[scale=0.55]{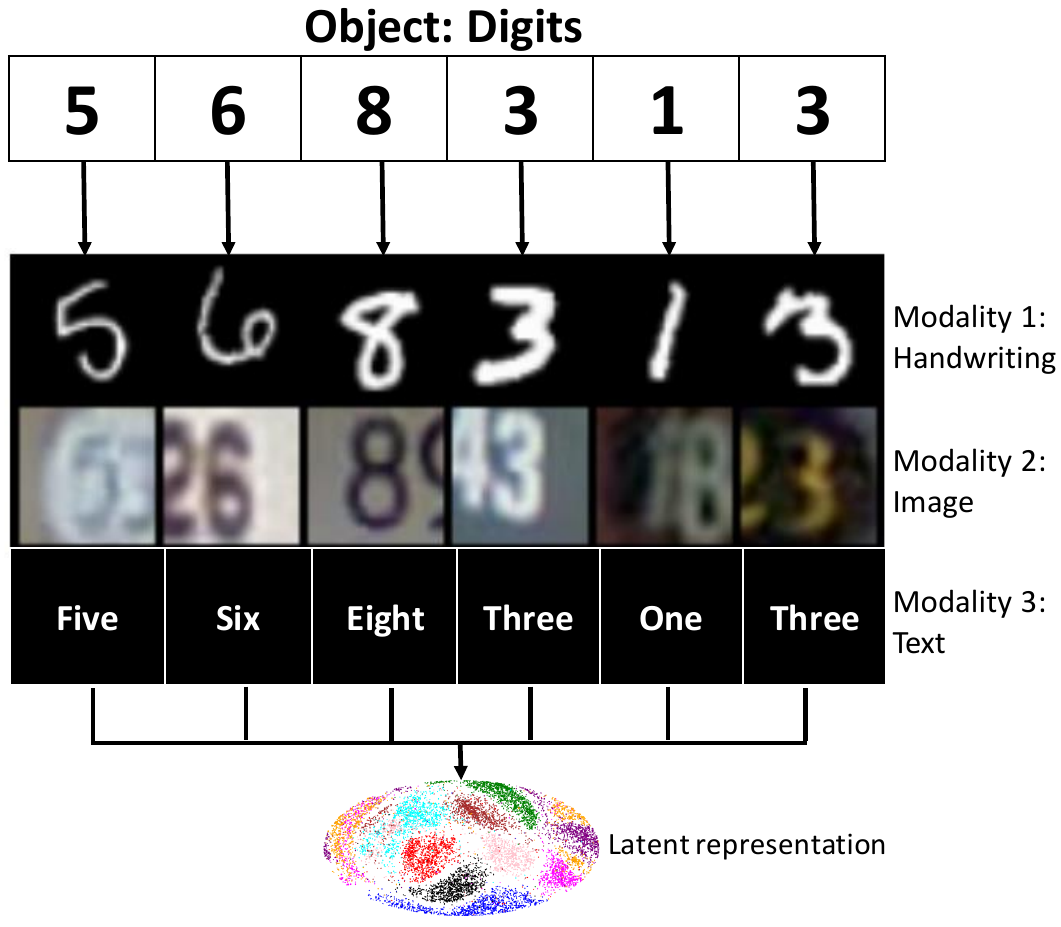}
    \caption{A graphical framework for multimodal learning in which we have access to 3 data modalities: handwriting, images, and text, describing the same object, a digit. Multimodal learning models relate these sources of information to learn a data representation that embeds information on all data modalities.}
    \label{fig_multimodal_rep}
\end{figure}
Multimodal learning is the field in machine learning that designs models which can process and relate information from different stimuli or data modalities \citep{baltruvsaitis2018multimodal}. There are two main approaches for multimodal representation learning: i) joint representation and ii) coordinated representation. The former uses all modalities as inputs and projects them into a common space, while the latter assumes that each representation exists in its own space and all representations are coordinated through a similarity or structure constraint \citep{baltruvsaitis2018multimodal}. However, it is common to use the term ``joint representations" to refer to data representations that embed information from multiple data modalities, regardless of the learning approach. Figure \ref{fig_multimodal_rep} \citep{mancisidor2021discriminative} shows a multimodal learning scheme, in which we have access to three different modalities (handwriting, image, and text) representing a common object (digits). A multimodal learning model learns a data representation that embeds useful information from each data modality. 

In this research, we use the CMMD model, which uses the joint representation approach for representation learning and at the same time trains a classifier conditioned on the learned latent representations. In addition, as we explain in Section \ref{sec_method_cmmd}, the CMMD model optimizes mutual information between missing modalities and latent representations used for classification, achieving a more accurate classification.

\subsection{Variational Inference}
The objective function optimized by the CMMD model is obtained by using the Variational Inference (VI) approach. Therefore, in what follows, we provide a short description of this method. Assume that we have access to $N$ observations and latent variables, i.e., $\bm{X}=\{\x_1,\x_2,\cdots,\x_N\}$ and $\bm{Z}=\{\z_1,\z_2,\cdots,\z_N\}$, that are related through the joint density $p(\bm{Z},\bm{X})=p(\bm{X}|\bm{Z})P(\bm{Z})$. Assuming a full factorization for $p(\bm{Z})=\prod_{n=1}^N p(\z_i)$ and $p(\bm{X}|\bm{Z})=\prod_{n=1}^N p(\x_i|\z_i)$, the average marginal log-likelihood of the data set $\bm{X}$ is simply $\frac{1}{N} \log p(\bm{X}) =\frac{1}{N} \sum_{i=1}^N\log p(\x_i)$. The marginal log-likelihood is, however, intractable due to the integral $\int p(\bm{X},\bm{Z})d\bm{Z}$ \citep{kingma2013auto,blei2017variational}. VI circumvents this problem by noting that the marginal log-likelihood for the \textit{i}th observation can be written as 
\begin{align}
    \log p(\x^i)  =& \log \int p(\x^i,\z)d\z \nonumber \\
                =& \log \int q(\z|\x^i) \frac{p(\x^i|\z)p(\z)}{q(\z|\x^i)}d\z \nonumber \\
                =& \log \mathbb{E}_{q(\z|\x^i)}\frac{p(\x^i|\z)p(\z)}{q(\z|\x^i)} \nonumber \\
                \geq& \mathbb{E}_{q(\z|\x^i)} [\log p(\x^i|\z) + \log p(\z) - \log q(\z|\x^i)] \nonumber \\
                \equiv& \text{ELBO}
\label{eq_elbo}
\end{align}
where $q(\z|\x)$ is a variational distribution (also called inference or recognition model) approximating the true posterior distribution $p(\z|\x)$, and the inequality is a result of the concavity of log and Jensen's inequality. Hence, VI optimizes intractable problems by introducing variational distributions and maximizing the \textit{Evidence Lower Bound} (ELBO) in Equation \ref{eq_elbo}, instead of the intractable marginal log-likelihood. Note that the ELBO in Equation \ref{eq_elbo} can be derived by minimizing the divergence $KL[q(\z|\x)||p(\z|\x)]$ and, in that case, the ELBO can be rewritten as $\text{ELBO} = \log p(\x^i) - KL[q(\z|\x^i)||p(\z|\x^i)]$. Therefore, maximizing the ELBO is equivalent to minimizing the Kullbak-Leibler divergence between the true posterior and its variational approximation, which turns out to improve the tightness of the lower bound.

\subsubsection{Variational Inference with Deep Neural Networks}
\begin{figure}
    \centering
    \includegraphics[scale=1]{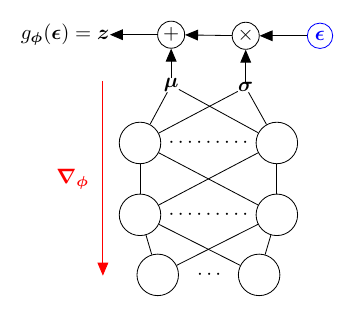}
    \caption{Architecture of a neural network that learns the parameters $\bm{\mu}$ and $\bm{\sigma}$ of a Gaussian distribution. The output layer $\bm{\mu}$ has the same number of neurons as the dimensionality of $\z$. Then we draw a representation $\z$ by using the location-scale transformation  $\z = \bm{\mu} + \bm{\sigma} \odot \bm{\epsilon}$, where $\bm{\epsilon} \sim \mathcal{N}(\bm{0},\bm{1})$ and $\odot$ is the element-wise product.}
    \label{fig_reptrick}
\end{figure}

\cite{kingma2013auto} and \cite{rezende2014stochastic} coupled VI methods with deep neural networks and show how these models can be efficiently optimized by backpropagation and the  AutoEncoding Variational Bayesian (AEVB) algorithm \citep{kingma2013auto}. Assume that both densities $q(\z|\x)$ and $p(\x|\z)$ in Equation \ref{eq_elbo} follow a Gaussian distribution with diagonal covariance matrix, and that $p(\z)$ is an isotropic Gaussian distribution, where $\z \in \mathbb{R}^d$. The ELBO\footnote{From now on we drop the superscript \textit{i} in order to not clutter notation.} in Equation \ref{eq_elbo} then takes the form
\begin{align*}
    \text{ELBO} =& \log p_{\bm{\theta}}(\x|\z) - KL[q_{\bm{\phi}}(\z|\x)||p(\z)] \nonumber \\
                =& \log \mathcal{N}(\x|\z; \bm{\mu}=f_{\theta}(\z), \bm{\sigma}^2=f_{\theta}(\z)) - KL[\mathcal{N}(\z|\x; \bm{\mu}=f_{\phi}(\x), \bm{\sigma}^2=f_{\phi}(\x))||\mathcal{N}(\z; \bm{0}, \bm{1})] \nonumber \\
                =& \log \mathcal{N}(\x|\z; \bm{\mu}=f_{\theta}(\z), \bm{\sigma}^2=f_{\theta}(\z)) + \frac{1}{2}\sum_{j=1}^d(1+\log \sigma_j^2-\mu_j^2 - \sigma_j^2),
\end{align*}
where $f(\cdot)$ denotes a function composed by a neural network with trainable parameters $\bm{\theta}$ and $\bm{\phi}$ for the generative and inference model, respectively, and $d$ is the dimension of $\z$. Given that the variational distribution and the prior density are both Gaussian, the Kullback-Leibler divergence has a closed-form. Figure \ref{fig_reptrick} shows the architecture of a neural network for the inference model $q(\z|\x)$. Note that the output layer of such a neural network contains the Gaussian parameters, and that $\z$ is drawn using the location-scale transformation  $\z = \bm{\mu} + \bm{\sigma} \odot \bm{\epsilon}$ where $\bm{\epsilon} \sim \mathcal{N}(\bm{0},\bm{1})$ and $\odot$ is the element-wise product. Such an architecture has the advantage that it can be backpropagated. This implies that by updating the trainable parameters $\bm{\theta}$ and $\bm{\phi}$ of the neural network, we learn the parameters $\bm{\mu}$ and $\bm{\sigma}$ for the inference and generative model, respectively, that maximize the ELBO. 

At this point, it is worth mentioning that the inference model $q(\z|\x)$ generates a code of the input data and the generative model $p(\x|\z)$ takes that code and generates a new instance of the input data. Hence, $q(\z|\x)$ is often referred to as a probabilistic encoder and $p(\x|\z)$ is referred to as a probabilistic decoder. Further, note that the code $\z$ is just a representation in a latent space for the input data $\x$. Therefore, it is common to call $\z$ a \textit{data representation} or simply a \textit{representation} in short.

\subsection{Conditional MultiModal Discriminative Model (CMMD)}\label{sec_method_cmmd}
The CMMD model relates information from multiple data modalities, assuming that we have access to all modalities and class labels only for model training. That is, at training time we observe $(\xo,\xm,y)$, where $\xo=(\x_1,\cdots,\x_n)$ are \textit{n} modalities that are always observed, and $\xm=(\x_{n+1},\cdots,\x_{n+m})$ are \textit{m} modalities that together with class labels $y$ are missing at test time. Hence, only $\xo$ is available during both training and test time. At test time, the CMMD model generates multimodal representations $\z$ using a prior distribution $p(\z|\xo)$ conditioned on the observed modalities. These representations $\z \sim p(\z|\xo)$ are used in both the generative process $p(\xm|\xo,\z)$ and in the classifier model $p(y|\z)$. This learning process encourages the CMMD model to learn multimodal representations that are useful for classification and generating the missing modalities at test time.

\begin{figure}[!t]
    \centering
    \includegraphics[scale=0.8]{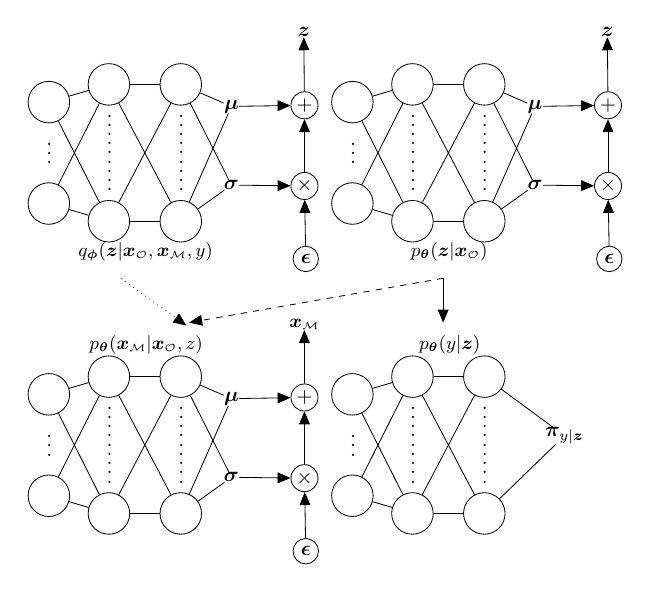}
    \caption{Architecture for the CMMD model that is composed by 4 neural networks: encoder, decoder, prior, and classifier. The dotted arrow indicates a forward pass during training, which is replaced by the dashed arrow at test time, i.e., the input to $p(\xm|\xo,\z)$ is $\z \sim q(\z|\xo,\xm,y)$ during training, while  $\z \sim p(\z|\xo)$ at test time. The solid arrow depicts a common forward propagation during training and testing, i.e., the input to $p(y|\z)$ is always $\z \sim p(\z|\xo)$.}
    \label{fig_cmmd}
\end{figure}

The generative model in CMMD factorizes as $p(\xm,y,\z|\xo) = p(y|\z)p(\z|\xo)p(\xm|\xo,\z)$ and, under this model specification, the posterior distribution $p(\z|\xo,\xm,y)$ is intractable \citep{mancisidor2021discriminative}. Therefore, CMMD uses VI and approximates the true posterior distribution with a variational density $q(\z|\xo,\xm,y)$. Hence, the variational lower bound on the marginal log-likelihood of the data is
\begin{align}
\label{cmmdelbo}    
    \log p(\xm,y|\xo) =& \log \int p(\xm,y,\z|\xo) d\z \nonumber \\
                      =& \log \int q(\z|\xo,\xm,y)\frac{p(\xm,y,\z|\xo)}{q(\z|\xo,\xm,y)} d\z \nonumber \\
                      =& \log \mathbb{E}_{q(\z|\xo,\xm,y)} \frac{p(\xm,y,\z|\xo)}{q(\z|\xo,\xm,y)} \nonumber \\
                      \geq&  \mathbb{E}_{q(\z|\xo,\xm,y)}\bigg[\log \frac{p(\xm,y,\z|\xo)}{q(\z|\xo,\xm,y)}\bigg] \nonumber \\
                      =& \mathbb{E}_{q(\z|\xo,\xm,y)}[\log p(\xm|\xo,\z) + \log p(y|\z) + \log p(\z|\xo) - \log q(\z|\xo,\xm,y)]\nonumber \\
                      =& \mathbb{E}_{q(\z|\xo,\xm,y)}[\log p(\xm|\xo,\z) + \log p(y|\z)] - KL[q(\z|\xo,\xm,y) || \log p(\z|\xo) ],
\end{align}
where the inequality is a result of the concavity of log and Jensen's inequality. Equation \ref{cmmdelbo} contains an upper bound on the mutual information between $\xm$ and $\z$ \citep{mancisidor2021discriminative}, which is exactly a property that we want to maximize. That is, we are interested in learning representations $\z$ that embed as much information as possible about the missing modalities $\xm$. To achieve this, the CMMD model includes a conditional mutual information $I(\xm,\z|\xo)$ into the lower bound in Equation \ref{cmmdelbo} to obtain the following \textit{likelihood-free} objective function\footnote{The reader is referred to \cite{mancisidor2021discriminative} for a complete derivation of the objective function.}
\begin{align}
    \mathcal{J}(\xo,\xm,y) =& \mathbb{E}_{q(\z|\xo,\xm,y)}[\log p(\xm|\xo,\z) + \log p(y|\z)] \nonumber \\
    -& KL[q(\z|\xo,\xm,y) || \log p(\z|\xo) ] + (1-\omega)I(\xm,\z|\xo) \nonumber \\
    =& \mathbb{E}_{q(\z|\xo,\xm,y)}[\log p(\xm|\xo,\z) + \log p(y|\z)] \nonumber \\
    -& \omega KL[q(\z|\xo,\xm,y)||p(\z|\xo)] - (1-\omega)KL[q(\z|\xo)||p(\z|\xo)],
\label{eq_cmmdobjfun}
\end{align}
where $\omega \in [0,1]$ is a weight hyperparameter controlling the influence of the mutual information term. Note that for $\omega = 1$, Equation \ref{eq_cmmdobjfun} recovers the evidence lower bound in Equation \ref{cmmdelbo}. The density functions in the CMMD model are assumed to be
\begin{align}
    p(\xm|\xo,\z) \sim& \mathcal{N}(\xm|\xo,\z; \bm{\mu}=f_{\bm{\theta}}(\xo,\z),\bm{\sigma}^2=f_{\bm{\theta}}(\xo,\z)) \nonumber \\
    p(\z|\xo) \sim& \mathcal{N}(\z|\xo; \bm{\mu}=f_{\bm{\theta}}(\xo),\bm{\sigma}^2=f_{\bm{\theta}}(\xo)) \nonumber \\
    q(\z|\xo,\xm,y) \sim& \mathcal{N}(\z|\xo,\xm,y; \bm{\mu}=f_{\bm{\phi}}(\xo,\xm,y),\bm{\sigma}^2=f_{\bm{\phi}}(\xo,\xm,y)) \nonumber \\
    p(y|\z) \sim& \text{Bernoulli}(y|\z; \pi=f_{\bm{\theta}}(\z)), 
    \label{eq_cmmddensities}
\end{align}
where Gaussian distributions $\mathcal{N}(\cdot)$ assume a diagonal covariance matrix, $f(\cdot)$ denotes a multilayer perceptron model that learns density parameters, $\bm{\theta}$ and $\bm{\phi}$ denote all trainable neural network weights for the generative and inference model, respectively, and ($\bm{\mu}$, $\bm{\sigma}^2$) and $\pi$ are the density parameters for the Gaussian and Bernoulli distributions. Figure \ref{fig_cmmd} \citep{mancisidor2021discriminative} shows the architecture for the CMMD model, which contains a posterior $q(\z|\xo,\xm,y)$ and a prior $p(\z|\xo)$ distributions for multimodal representations, a generative model $p(\xm|\xo,\z)$ for missing modalities, and a classifier $p(y|\z)$ for class labels.

It is noteworthy that minimizing the divergence term $KL[q(\z|\xo,\xm,y)||p(\z|\xo)]$ in Equation \ref{eq_cmmdobjfun} has the effect of bringing the prior close to the posterior \citep{suzuki2016joint} and it also minimizes the expected information required to convert a sample from the prior into a sample from the posterior distribution \citep{doersch2016tutorial}. This effect is critical to learn multimodal representations that embed information from all data modalities.

\subsubsection{CMMD for Corporate Bankruptcy Prediction} 
In the context of corporate bankruptcy prediction\footnote{This research focuses on public companies in the AMEX, NYSE, and NASDAQ stock markets.}, the modalities that are always observed are $\xo = (\x_1,\x_2)$, representing accounting and market information, since this information can easily be obtained for all companies on quarterly and daily basis, respectively. On the other hand, $\xm=\x_3$ and $y$ correspond to the MDA section in Form 10-K and class labels, respectively, which are assumed to be missing at test time. It makes sense to treat the MDA section as a modality that is not always observable, due to the fact that we rely on technical procedures to extract it from Form 10-K. In addition, not all public companies are under obligation to file annual reports containing such a section. \review{Note that the distribution $p(\xm|\xo,\z)$ can be used to generate the missing MDA modality, conditioned on the available modalities $\xo$ and on latent representations drawn from the prior distribution $p(\z|\xo)$.} Furthermore, in a real forecast scenario we do not have access to class labels and therefore they cannot be treated as an always observable modality.
%On the other hand, both accounting and market data can be easily obtained for all companies on a quarterly and daily basis, respectively.

The CMMD model uses accounting and market modalities to define an informative prior distribution, which generates data representations. When the MDA is available during model training, the CMMD model draws data representations from a posterior distribution that is updated by this data modality. Furthermore, class posterior probabilities $\pi$ in the CMMD classifier are estimated based on data representations, i.e., $p(y|\z)$ with $\z \sim p(\z|\xo)$. This classification approach differs from the traditional method where posterior probabilities $\pi'$ are estimated based on observed features, i.e., $p(y|\xo)$. Through an efficient learning mechanism\footnote{See the discussion at the end of Section \ref{sec_method_cmmd}.}, the CMMD model aligns the parameters in the prior and posterior distributions. Therefore, the CMMD classifier anchors its forecasts on data representations that, despite being drawn by the prior distribution, contain information from all data modalities. 

To make this point clear, let $q(y|\xo) \sim \text{Bernoulli}(\pi')$ be a classifier model that estimates posterior class probabilities $\pi'$ given $\xo$, and the let the CMMD classifier be specified as in Equation \ref{eq_cmmddensities}. Further, let $\bm{Z}_{{\Xo}_{te}}$ denote a data set of representations drawn from the CMMD prior distribution $p(\z|\xo)$, where ${\Xo}_{te}$ is a test data set composed by the observable modalities $\xo$. Given that the CMMD model estimates class posterior probabilities based on representations that embed information from all data modalities, the classification performance of posterior probabilities $\pi_i$ for all $i$ in $\bm{Z}_{{\Xo}_{te}}$, should be relatively higher than that of posterior probabilities $\pi'_j$ for all $j$ in ${\Xo}_{te}$.

\section{Data}\label{sec_data}
The data used in this research represent the largest data set ever used for bankruptcy prediction (Table \ref{tbl_overview}). The data set includes accounting, market, and MDA data for publicly traded firms in NYSE, NASDAQ, and AMEX stock exchanges in the period 1994-2020. Accounting and market data are extracted from Wharton Research Data Services (WRDS), which provides access to Compustat Fundamentals and to the Center for Research in Security Prices (CRSP). The MDA section is directly extracted from the Form 10-K, which is available at the Electronic Data, Gathering, Analysis, and Retrieval (EDGAR) system provided by the U.S. Securities and Exchange Commission (SEC)\footnote{Form 10-K was for the first time publicly available through EDGAR in 1994/1995.}. The bankruptcy data used in this research are an updated version of the data used in \cite{chava2014environmental}, which are a comprehensive sample including the majority of publicly listed firms that filed for either Chapter 7 or 11 in the period 1964-2020.

\begin{table}[t]
\scriptsize
\centering
\caption{Accounting and market variables used for bankruptcy prediction. The first column shows the variable names, while the second column shows their corresponding column names in the WRDS data base. All accounting information is extracted from \texttt{comp.fundq} and market information from \texttt{crsp.msf} and \texttt{crsp.dsf}. In the column Variable, we use the following abbreviations: ME\_TL is market equity + total liabilities, TE is total equity and is calculated as the equity sum for all companies in a given quarter, C\&SI is Cash and Short-term Investment, RE is Retained Earnings, CL is Current Liabilities}
\def\arraystretch{1.1}
\setlength{\tabcolsep}{3.5pt}
\begin{tabular}{|ll|ll|}
\hline
Variable &Database Names&Variable &Database Names\\
\hline
1) Assets/Liabilities&actq/lctq&18) Market-to-Book Ratio&(cshoq*prccq)/(atq-ltq)\\
2) Accounts payable/Sales&apq/saleq&19) Net Income/Total Assets&niq/atq\\
3) C\&SI/Total Assets&cheq/atq&20) Net Income/ME\_TL&niq/(cshoq*prccq+ltq)\\
4) C\&SI/ME\_TL&cheq/(cshoq*prccq+ltq)&21) Net Income/Sales&niq/saleq\\
5) Cash/Total Assets&chq/atq&22) Operating Income/Total Asset&oiadpq/atq\\
6) Cash/CL&chq/lctq&23) Operating Income/Sales&oiadpq/saleq\\
7) Total Debts/Total Assets&(dlcq+0.5*dlttq)/atq&24) Quick Assets/CL&(actq-invtq)/lctq\\
8) Growth of Inventories /Inventories&invchy/saley&25) RE/Total Asset&req/atq\\
9) Inventories/Sales&invtq/saleq&26) RE/CL&req/lctq\\
10) (CL – Cash)/Total Asset&(lctq-chq)/atq&27) Sales/Total Assets&saleq/atq\\
11) CL/Total Asset&lctq/atq&28) Equity/Total Asset&seqq/atq\\
12) CL/Total Liabilities&lctq/ltq&29) Working Capital/Total Assets&wcapq/atq\\
13) CL/Sales&lctq/saleq&30) Rsize&log(cshoq*prccq)/TE\\
14) Total Liabilities/Total Assets&ltq/atq&31) Log Price&log(min(prccq,15))\\
15) Total Liabilities/ME\_TL&ltq/(cshoq*prccq+ltq)&32) Excess Return Over S\&P 500& ret - vwretd\\
16) Log(Total Assets)&log(atq)&33) Stock Volatility&std(ret)\\
17) Log(Sale)&log(abs(saleq))&&\\
\hline
\end{tabular}
\label{tbl_variables}
\end{table}

\begin{figure}[!t]
    \begin{subfigure}{0.45\textwidth}
    \includegraphics[scale=0.4]{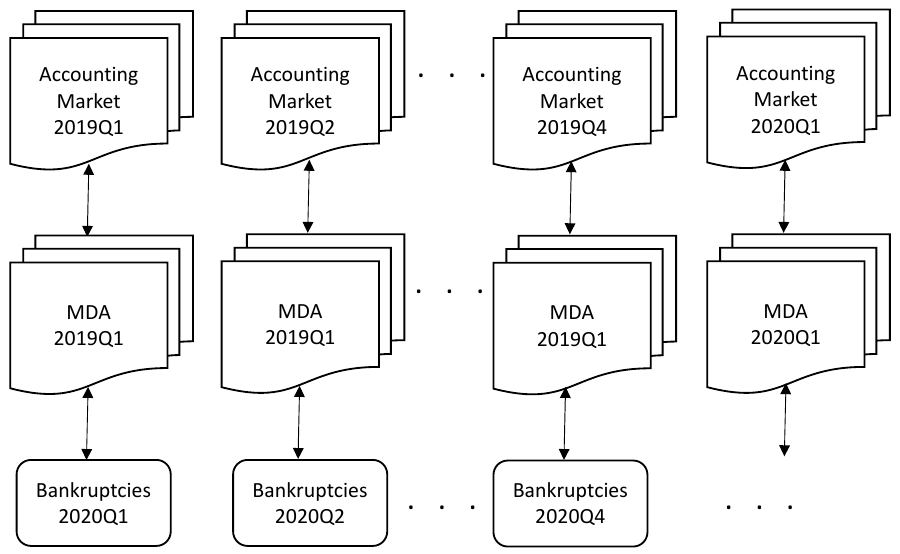}
    \caption{}
    \end{subfigure}
    \hspace{-1.1cm}
    \begin{subfigure}{0.55\textwidth}
    \includegraphics[scale=0.52]{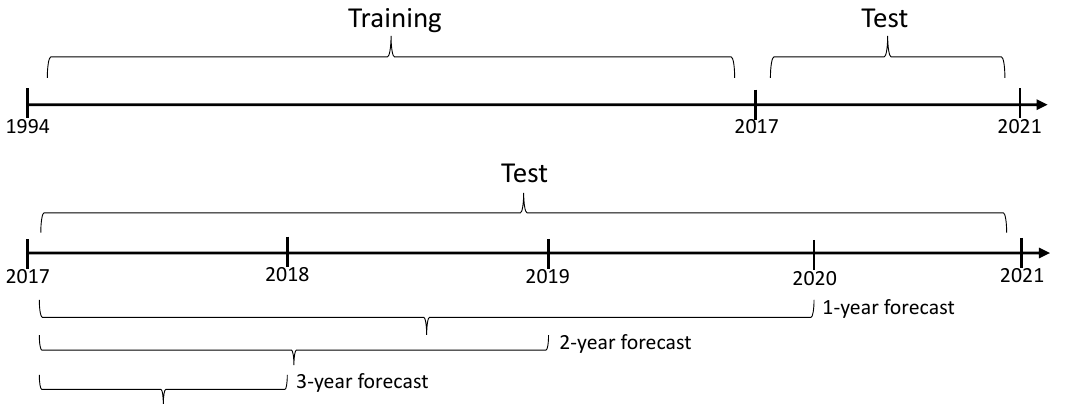}
    \caption{}
    \end{subfigure}
    \caption{(a): We collect 31 accounting and 2 market predictors that are merged with an MDA from the same quarter. We roll-over the MDA data for 3 more quarters and merge it with the corresponding predictors in those 3 quarters, i.e., an MDA can be used during 4 quarters. (b): The data set in this research includes bankruptcies from 1994 to 2020. We use data from 1994 to 2016 for model training, while data from 2017 to 2020 are used for testing the predictive power of the models. For 1-year forecasts, for example, the latest quarter in which we can make a forecast is 2019Q4, which corresponds to bankruptcies during 2020Q4.}
    \label{fig_dataprocess}
\end{figure}

Our data set is constructed in a similar way as in \cite{shumway_forecasting_2001} and \cite{mai_deep_2019}, i.e., letting each firm-quarter represents a separate observation. Hence, for each quarter we collect 31 accounting and 2 market predictors (Table \ref{tbl_variables}). These 33 predictors are merged with the MDA corresponding to the same quarter. To construct a quarterly panel data, we roll-over the MDA data for 3 more quarters merging it with the corresponding predictors in the following 3 quarters. Hence, the same MDA is used for 4 quarters (Figure \ref{fig_dataprocess} panel (a)). After 4 quarters, we should get a new MDA that can be merged with its corresponding 33 predictors. If that is not the case, all firm-quarter observations will be missing until we get a new MDA. Finally, to make sure that all information is available for forecasting, we lag all observations by one period. 

\begin{figure}[!t]
    \centering
    \includegraphics[scale=0.5]{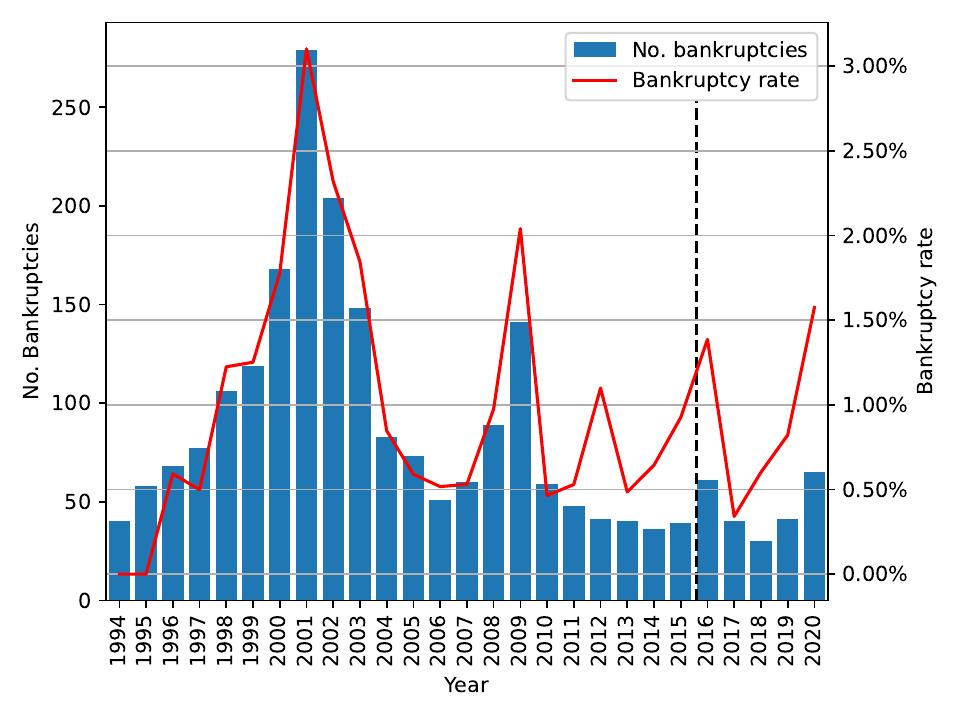}
    \caption{Number of yearly bankruptcies and bankruptcy rates in our data set for 1-year bankruptcy predictions. Bankruptcy rate includes firm-quarter observations for which we could either merge the 3 data modalities or merge the 3 data modalities with 1-year ahead bankruptcies, while the number of bankruptcies include all listed companies that file for Chapter 7 or 11.}
    \label{fig_bankrate}
\end{figure}

In this research, we predict bankruptcy for three different forecast horizons: 1, 2, and 3 years (Figure \ref{fig_dataprocess} panel (b)). Therefore, \review{at test time,} the set of 33 predictors is used to predict 1, 2, and 3 years-ahead bankruptcies (Figure \ref{fig_dataprocess} panel (a)), \review{e.g., assuming a 1-year forecast horizon, all predictive variables available in 2019Q4 are used to predict bankruptcies in 2020Q4}. For simplicity, we remove all firms from our data set after they file either for Chapter 7 or 11, i.e., the data set does not include reorganized firms after Chapter 11 was filed. Bankruptcies for which we cannot merge a set of predictors, regardless of the forecasting horizons, are discarded from our data set. After merging accounting, market, and MDA data, our data set for 1-year predictions contains 181,472 observations for model training, of which 699 are bankruptcies, and 40,950 observations for testing the model performance, of which 110 are bankruptcies. For 2-years predictions, the training data has 179,181 observations, of which 668 are bankruptcies, and the test data has 27,369 observations with 84 bankruptcies. Finally, the data set for 3-year predictions has 177,835 observations for model training, containing 588 bankruptcies, and 13,725 observations for testing the model, of which 54 are bankruptcies. Figure \ref{fig_bankrate} shows the number of yearly bankruptcies and bankruptcy rates in our data set for 1-year bankruptcy predictions. Note that bankruptcy rates include only firm-quarter observations for which we could either merge the 3 data modalities (healthy firms) or the 3 data modalities with the bankruptcy data  (bankrupted firms). On the other hand, the number of bankruptcies represents all yearly bankrupted firms in our bankruptcy data. This difference explains the discrepancy between bankruptcy rate and number of bankruptcies in 2012, as shown in Figure \ref{fig_bankrate}. 

\subsection{Data preprocessing}
MDA sections are extracted from the company's annual reports in Form 10-K and are transformed into a clean text file. Then, we follow the basic steps in natural language processing, i.e., word tokenization, remove stopwords, and stemming. We observe that some MDA documents do not contain substantial information, hence we include MDA documents containing more than 1,500 word tokens. This ensures that MDA documents contain substantial information. We use Term Frequency-Inverse Document Frequency (TF-IDF) to convert the preprocessed  MDA documents into a numerical representation with 20,000 dimensions, which is the same number of dimensions as in \cite{mai_deep_2019}. Note that the transformation of MDA into TF-IDF allows us to generate tokens of this modality with the CMMD model (see Section \ref{sec_generative}).  

We use similar accounting ratios as in previous research, e.g., \cite{altman_financial_1968,beaver1966financial} and \cite{mai_deep_2019}, which reflect the liability, liquidity, and profitability for each company. The main difference with our data set is, however, that we construct a panel data with quarterly observations as shown in Figure \ref{fig_dataprocess} panel (a). By doing that, our data set is significantly larger than previous data sets, which is needed to train a model like CMMD. The data set used in \cite{mai_deep_2019}, for example, contains only 99,994 firm-year observations compared to 181,472 firm-quarter observations in our data set. All accounting variables are extracted from the Compustat table \texttt{comp.funq} and their calculation is shown in Table \ref{tbl_variables}. In addition, excess returns, defined as the stock return relative to the S\&P 500 returns, and stock volatility, defined as the standard deviation for stock returns in the past 63 days,  are extracted from CRSP tables \texttt{crsp.msf} and \texttt{crsp.dsf}. Finally, accounting and market variables are scaled in the range 0 and 1 to match the scale of TF-IDF vectors. 

Given that the class labels in the data set are highly imbalanced, we down-sample the majority class ($y=0$) until it equals the number of observations in the minority class ($y=1$) for model training. On the other hand, the test data set preserves its original number of classes, ensuring that models are tested on real scenarios.  

\section{Experiments and Results}\label{sec_experiments}
This section compares our proposed methodology for bankruptcy prediction, which is based on multimodal representations generated by the CMMD model, with the traditional approach in which the classifier model is trained and tested on the same observed predictors. To that end, we include the following benchmark models: Logistic Regression (LR), Support Vector Machine (SVM), Multilayer Perceptron (MLP), k-Nearest Neighbors (k-NN), Random Forests (RF), and Naive Bayes (NB). We measure the classification performance of all models using different indicators that evaluate performance from different perspectives. Specifically, we use the Area Under the ROC Curve (AUC) and H-measure \citep{hand2009measuring} as global performance metrics, and the Kolmogorov-Smirnov (KS) test as a local performance metric. Note that all three metrics estimate classification performance in the interval [0 1]. \review{Furthermore, in sections \ref{sec_interpretability} and \ref{sec_generative} we use Shapley additive explanations to interpret the decisions made by our proposed methodology and test the generative capabilities of the CMMD model, respectively.} Finally, Section \ref{sec_representations} analyzes the information embedded in multimodal representations $\z \sim p(\z|\xo)$.

The CMMD model is implemented in TensorFlow\footnote{Code available at \href{https://github.com/rogelioamancisidor/bankruptcy}{https://github.com/rogelioamancisidor/bankruptcy}} and it is trained using the Adam optimizer \citep{kingma2014adam} with default values. To provide a fair model comparison, we fine-tune the hyperparameters of all models and for all forecast horizons. Appendix \ref{app_grid} shows the grid-search values in our fine-tuning approach, as well as the final architecture for each model and each forecast horizon.

\subsection{Experimental Design}\label{sec_expdesign}
We conduct three different experiments to compare the classification performance of our proposed methodology with that of all benchmark models. In Experiments I and II we use the time period 1994 - 2007 for training and the time period 2008 - 2014 for testing, while in Experiment III we use the time period 1994 - 2016 for training and the time period 2017 - 2020 for testing.  The difference between Experiment I and Experiment II is that the former uses training and test sets consisting of $\xo$ only for the benchmark models, while in the latter we also use $\x_3$, the 20,000 TF-IDF variables. In both experiments, the CMMD model uses all three modalities for training, and only the accounting and market data for testing. 

The time period used in Experiments I and II is chosen to provide a fair and direct comparison with the \textit{DL-Embedding + DL-1 Layer} model introduced in \cite{mai_deep_2019}.  Experiment III is similar to Experiment I, except for the fact that the training and test periods are different. These two experiments 
study the classical scenario in which multimodal learning models assess how much information from missing modalities has been embedded into the multimodal representation, see e.g., \cite{shi2019variational,wang2016deep,sutter2021generalized} or \cite{mancisidor2021discriminative}. Table \ref{tbl_observations} summarizes the data used in the three experiments. As can be seen from this table, the number of testing observations reduces from 89,582 in Experiment I to 36,240 in Experiment II. This is due to the fact that MDA information is not available for 37\% of the test observations in Experiment I (the number of companies reduces from 5,203 to 3,132). 

\begin{table}[!t]
\footnotesize
\centering
    \caption{Number of observations and bankruptcies, and bankruptcy rates during the two different time periods studied in this research. Experiments I and II are as explained in Section \ref{sec_expdesign}.}
    \centering
    \def\arraystretch{1}
    \setlength{\tabcolsep}{18pt}
    \begin{tabular}{|cc|cc|c|}
    \hline
     &     & \multicolumn{2}{c|}{From 1994 to 2014} & From 1994 to 2020 \\
    %\hline
          &&  Experiment I& Experiment II & Experiment III\\
    \hline
     \multirow{3}{*}{Training}      &Observations&  122,916& 122,916 & 181,472 \\
                                    &Bankruptcies&  537& 537 & 699 \\
                                    &Bankruptcy rate&  0.4369\%& 0.4369\% & 0.3852\%\\
    \hline
    \multirow{3}{*}{Testing}        &Observations&  89,582& 36,240 & 40,950 \\
                                    &Bankruptcies&  179& 94 & 110\\
                                    &Bankruptcy rate&  0.1998\%& 0.2594\% & 0.2686\%\\
    \hline
    \end{tabular}
    \label{tbl_observations}
\end{table}

\subsection{Results}

\paragraph{Experiments I and II:} Table \ref{tbl_aucresults2} measures the classification performance, based on AUC, for 1 year bankruptcy predictions during the first time period from 1994 to 2014. In both Experiment I and II, RF is the model with highest AUC, followed by CMMD and SVM. Further, the benchmark models actually achieve higher classification performance in Experiment I than in Experiment II, meaning that the MDA data does not improve the prediction accuracy for these models. This is in agreement with \cite{mai_deep_2019}, where the only model with higher AUC when MDA is used is the one introduced by the authors.  Note also that all methods except k-NN and NB achieve higher AUC values than the DL-Embedding model introduced in \cite{mai_deep_2019}. This suggests that using panel data based on firm-quarter observations gives more accurate bankruptcy predictions.

The CMMD-model is not able to beat the RF-model in these experiments. We believe that this is due to the fact that it requires more training data than the competing methods. Hence, in Experiment III we have repeated Experiment I for a longer training period. 

\begin{table}[t!]
\footnotesize
\centering
\caption{Model performance for 1-year bankruptcy predictions. We report average and standard deviation values for 5 different randomly selected training data sets during the period 1994 to 2007. The test period, in this case, is from 2008 to 2014. Results marked with ${\dagger}$ are taken from \cite{mai_deep_2019}.}
\def\arraystretch{1}
\setlength{\tabcolsep}{18pt}
\begin{tabular}{|c|ccc|}
\hline
\multicolumn{4}{|c|}{\textbf{Experiment I - $\x_3$ is missing at test time}}            \\ 
\cline{1-4}
Model Name &  AUC & H-measure & KS \\
\hline
k-NN                &   0.8334 $\pm$ 0.0055 &  0.3633 $\pm$ 0.0122 & 0.5489 $\pm$ 0.0112 \\
NB                  &   0.6299 $\pm$ 0.0106 &  0.1719 $\pm$ 0.0084 & 0.2204 $\pm$ 0.0056 \\
LR                  &   0.8856 $\pm$ 0.0030 &  0.4643 $\pm$ 0.0040 & 0.6299 $\pm$ 0.0070 \\
SVM                 &   0.8868 $\pm$ 0.0037 &  0.4717 $\pm$ 0.0060 & 0.6289 $\pm$ 0.0123 \\
RF                  &   \textbf{0.9234} $\pm$ 0.0024  & \textbf{0.5636} $\pm$ 0.0073 & \textbf{0.7302} $\pm$ 0.0055 \\
MLP                 &   0.8738 $\pm$ 0.0091  & 0.4797 $\pm$ 0.0172 & 0.6251 $\pm$ 0.0176 \\
CMMD                &   0.8988 $\pm$ 0.0008  & 0.4847 $\pm$ 0.0020 & 0.6551 $\pm$ 0.0063  \\
\hline
\multicolumn{4}{|c|}{\textbf{Experiment II - $\x_3$ is available at test time}}            \\ 
\cline{1-4}
k-NN                &   0.8253 $\pm$ 0.0233 &  0.3544 $\pm$ 0.0563  &  0.5358 $\pm$ 0.0427 \\
NB                  &   0.6113 $\pm$ 0.0465 &  0.1305 $\pm$ 0.0654  & 0.1799 $\pm$ 0.0699 \\
LR                  &   0.8888 $\pm$ 0.0038 &  0.4792 $\pm$ 0.0034  & 0.6389 $\pm$ 0.0057 \\
SVM                 &   0.8963 $\pm$ 0.0045 &  0.5159 $\pm$ 0.0061 & 0.6538 $\pm$ 0.0094 \\
RF                  &   \textbf{0.9212} $\pm$ 0.0082 & \textbf{0.5675} $\pm$ 0.0275 & \textbf{0.7310} $\pm$ 0.0241 \\
MLP                 &   0.8773 $\pm$ 0.0052 &  0.4755 $\pm$ 0.0058 & 0.6241 $\pm$ 0.0103  \\
CMMD                &   0.8913 $\pm$ 0.0003 &  0.5014 $\pm$ 0.0020   &  0.6576 $\pm$ 0.0107   \\
DL-Embedding + DL-1 Layer  &  0.8420$^{\dagger}$ &  & \\
\hline
\end{tabular}
\label{tbl_aucresults2}
\end{table}

\paragraph{Experiment III:} Table \ref{tbl_aucresults} compares the classification performance for three different forecast horizons. We report the average performance, as well as the standard deviation of 5 different randomly selected training data sets. The test period is always the same and only varies depending on the forecast horizon.  

\begin{table}[!t]
\footnotesize
\centering
\caption{Model performance for 1-, 2-, and 3-year bankruptcy predictions. Experiment III: We report average and standard deviation values for 5 different randomly selected training data sets. The test period is always the same and it depends on the forecasting horizon. Training and test periods are shown in Figure \ref{fig_dataprocess} panel (b).}
\def\arraystretch{1}
\setlength{\tabcolsep}{26pt}
\begin{tabular}{|c|ccc|}
\hline
\multirow{2}{*}{Model Name}     & \multicolumn{3}{c|}{\textbf{AUC}}          \\ %\cline{2-4}
                                & 1 year    & 2 years   & 3 years   \\ 
\hline
% AUC
k-NN                & 0.8669 $\pm$ 0.0090   & 0.7809 $\pm$ 0.0088   & 0.7526 $\pm$ 0.0190  \\
NB                  & 0.8242 $\pm$ 0.0097   & 0.7328 $\pm$ 0.0085   & 0.7422 $\pm$ 0.0088  \\
LR                  & 0.8982 $\pm$ 0.0072   & 0.8229 $\pm$ 0.0061   & 0.8293 $\pm$ 0.0060  \\
SVM                 & 0.9017 $\pm$ 0.0057   & 0.8218 $\pm$ 0.0021   & 0.8167 $\pm$ 0.0023  \\
RF                  & 0.9251 $\pm$ 0.0047   & 0.8363 $\pm$ 0.0089   & 0.8404 $\pm$ 0.0092  \\
MLP                 & 0.9142 $\pm$ 0.0044   & 0.8355 $\pm$ 0.0083   & 0.8086 $\pm$ 0.0023  \\
CMMD                & $\bm{0.9315}$ $\pm$ 0.0011 &  $\bm{0.8581}$ $\pm$ 0.0031 & $\bm{0.8473}$ $\pm$ 0.0060 \\
%\hline
                                & \multicolumn{3}{c|}{\textbf{H-measure}}            \\ \cline{2-4}
                                %& 1 year    & 2 years   & 3 years           \\ 
\hline
% H
k-NN                & 0.4602 $\pm$ 0.0355   & 0.2472 $\pm$ 0.0077 & 0.1968 $\pm$ 0.0457  \\
NB                  & 0.4116 $\pm$ 0.0292   & 0.2022 $\pm$ 0.0071 & 0.2162 $\pm$ 0.0124  \\
LR                  & 0.5710 $\pm$ 0.0057   & 0.3573 $\pm$ 0.0101 & 0.3486 $\pm$ 0.0159  \\
SVM                 & 0.5764 $\pm$ 0.0040   & 0.3278 $\pm$ 0.0110 & 0.3177 $\pm$ 0.0014  \\
RF                  & 0.5938 $\pm$ 0.0071   & 0.3424 $\pm$ 0.0174 & \textbf{0.3944} $\pm$ 0.0342  \\
MLP                 & 0.5875 $\pm$ 0.0108   & 0.3685 $\pm$ 0.0091 & 0.3085 $\pm$ 0.0208  \\
CMMD                & $\bm{0.6228}$ $\pm$ 0.0028   & \textbf{0.4058} $\pm$ 0.0043  & 0.3764 $\pm$ 0.0151  \\
                                & \multicolumn{3}{c|}{\textbf{KS}}                   \\ \cline{2-4}
\hline
k-NN                & 0.6225 $\pm$ 0.0278   & 0.4463 $\pm$ 0.0227 & 0.3886 $\pm$ 0.0297   \\
NB                  & 0.6085 $\pm$ 0.0179   & 0.4252 $\pm$ 0.0105 & 0.4403 $\pm$ 0.0228   \\
LR                  & 0.7278 $\pm$ 0.0113   & 0.5619 $\pm$ 0.0067 & 0.5532 $\pm$ 0.0189   \\
SVM                 & 0.7229 $\pm$ 0.0067   & 0.5326 $\pm$ 0.0104 & 0.5516 $\pm$ 0.0027   \\
RF                  & 0.7371 $\pm$ 0.0063   & 0.5251 $\pm$ 0.0129 & 0.5604 $\pm$ 0.0334   \\
MLP                 & 0.7363 $\pm$ 0.0100   & 0.5565 $\pm$ 0.0058 & 0.5327 $\pm$ 0.0289   \\
CMMD                & $\bm{0.7695}$ $\pm$ 0.0039   & \textbf{0.5720} $\pm$ 0.0033   &  \textbf{0.5880} $\pm$ 0.0178     \\
\hline
\end{tabular}
\label{tbl_aucresults}
\end{table}
As can be seen from the table, the CMMD model now achieves the highest AUC and KS values, which implies that the discriminative power of CMMD is higher than that of all benchmark models. All models perform better in this experiment than in Experiment I. However, for the RF-model the AUC-value is only slightly higher. Interestingly, the AUC and H-measure do not agree on the model with the most accurate 3-year predictions. For this particular forecast horizon, CMMD has the highest AUC value, while RF has the highest H-measure. The KS test suggests, however, that there is a threshold where the CMMD model achieves the largest separation of the two class labels. Hence, from this experiment, it is clear that if the training data set is sufficiently large, the CMMD model outperforms all benchmark models, suggesting that multimodal representations $\z$ embed information from all data modalities and stand out as the best predictor for corporate bankruptcy.  

In Experiments I and III the CMMD model is used to compute predictions for all test observations. One might also consider a mixed approach in which the CMMD model is only used in the cases where the MDA data is missing and e.g. the RF model is used in the remaining cases.  

\subsection{Model Interpretability}\label{sec_interpretability}
The interpretability of advanced deep generative models is compromised to achieve better model performance. Thankfully, over the past decade, there has been a growing interest in developing methods to elucidate these sophisticated models. In their work, \cite{lundberg2017unified} propose the Shapley Additive Explanations (SHAP), a unified approach that can be used to interpret predictions made by any model. The SHAP value for a given feature is the average expected marginal contribution of this feature after all possible feature combinations have been considered. Hence, the SHAP value considers the effect the feature has by itself and in combination with the other features in the model. The utilization of SHAP values presents a user-friendly method to interpret models, offering valuable insights into comprehending bankruptcy predictions using the CMMD model. This information holds great value for financial regulators, investors, and practitioners. %\review{It is noteworthy that the SHAP values presented in this section should be taken with caution, as some of the $\xo$ variables are correlated (see Appendix \ref{app_corr}) and the underlying SHAP methodology assumes that the predictor variables are uncorrelated. Therefore, the ranking of the variables with the greatest influence on the predictions could be misleading.}

\begin{figure}[t!]
    \centering
    \begin{subfigure}{0.68\textwidth}
        \includegraphics[scale=0.5]{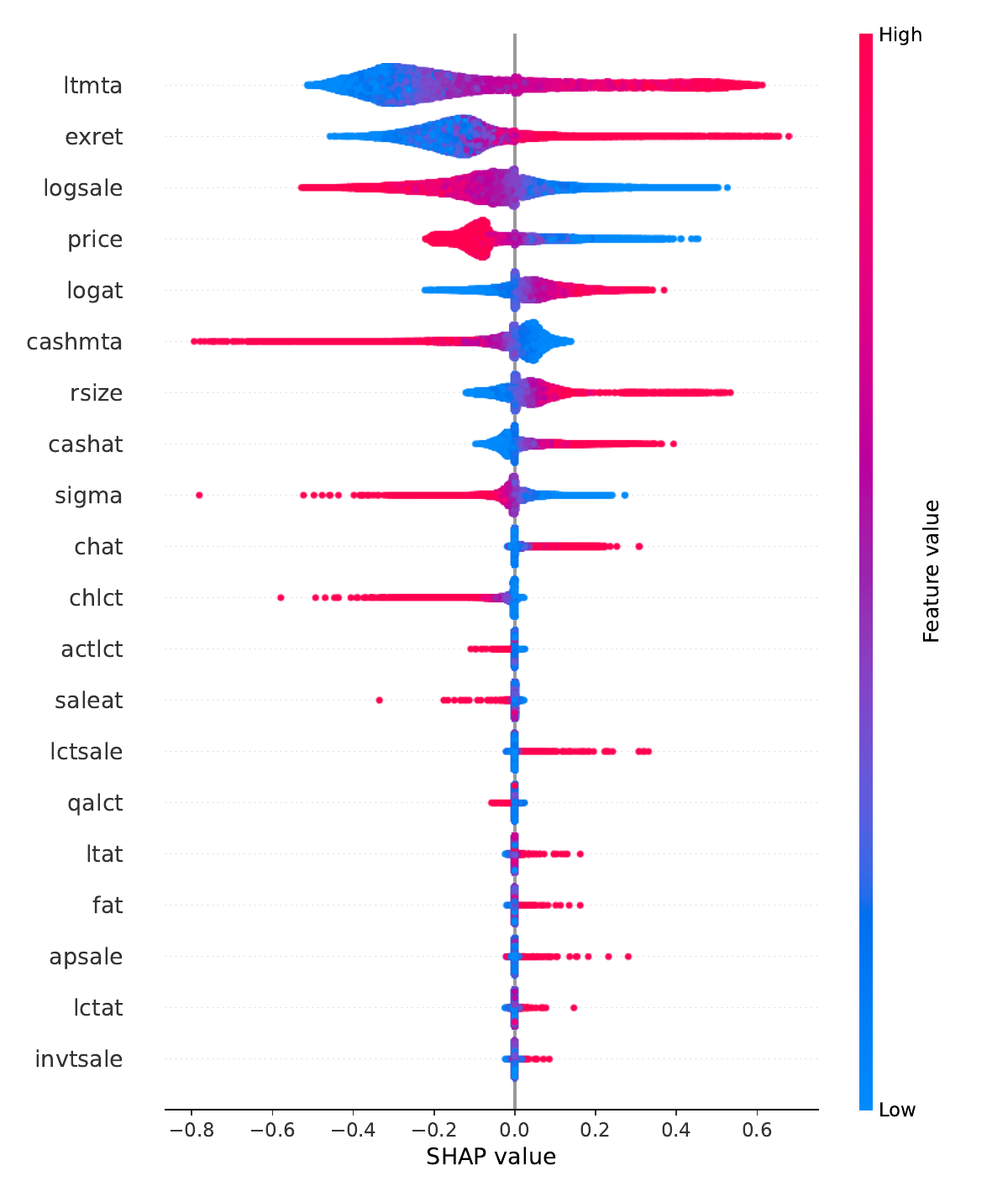}    
        \label{fig_shap_summary1}
    \end{subfigure}
    \hfill
    \begin{subfigure}{0.3\textwidth}
        \begin{subfigure}[t]{\textwidth}
            \includegraphics[scale=0.28]{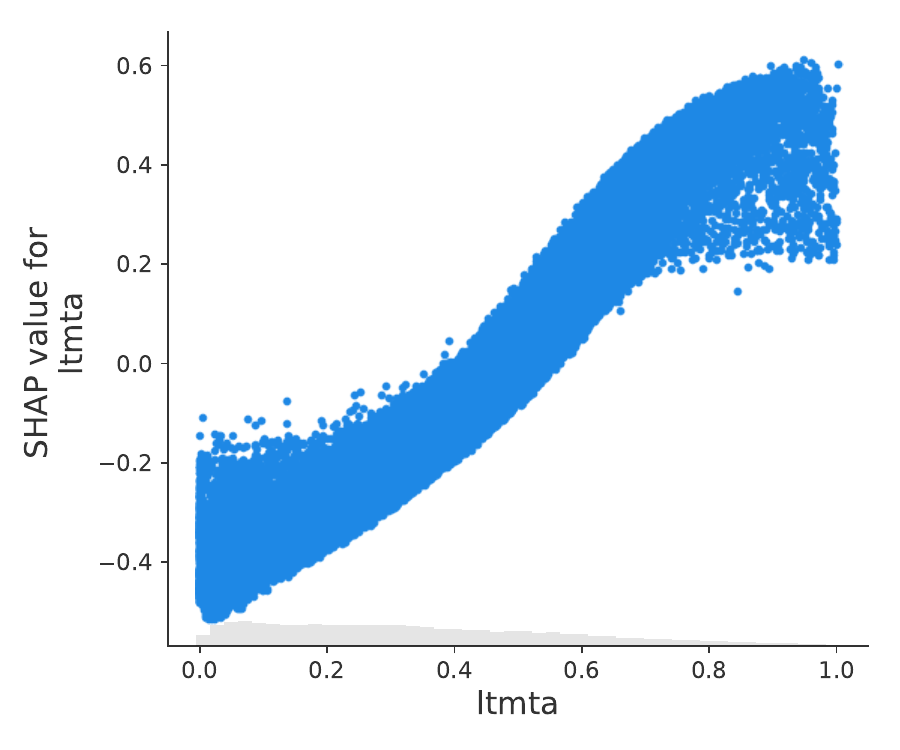}    
            \label{fig_shap_summary2}
        \end{subfigure}

        \begin{subfigure}{\textwidth}
            \includegraphics[scale=0.28]{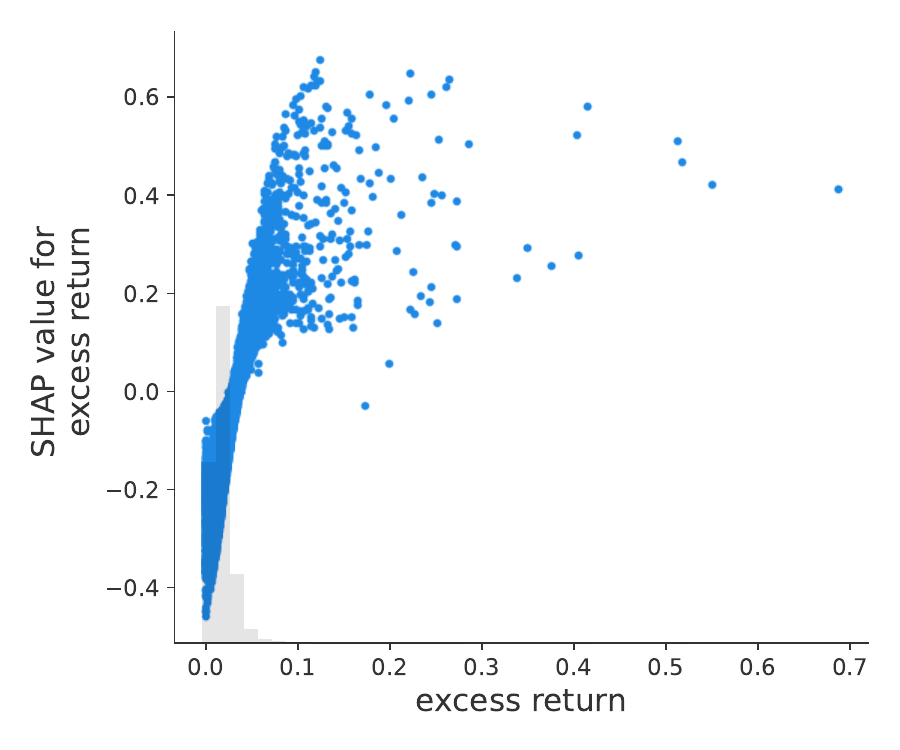}    
            \label{fig_shap_summary3}
        \end{subfigure}

        \begin{subfigure}[b]{\textwidth}
            \includegraphics[scale=0.28]{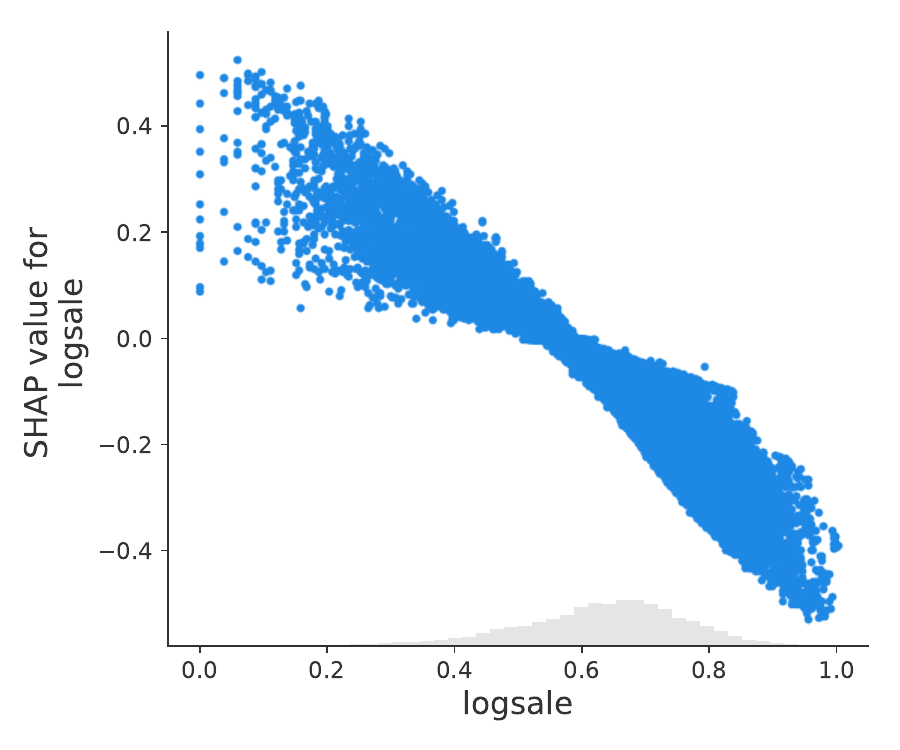}    
            \label{fig_shap_summary4}
        \end{subfigure}
    \end{subfigure}
    \caption{SHAP values for the 20 variables with the greatest impact on 1-year predictions (left panel). Figures on the right show the relationship between the three main variables with the greatest impact on 1-year bankruptcy predictions and their corresponding SHAP value. The variable names in Table \ref{tbl_variables} corresponding to the above variable codes are given in parentheses as follows: lmta(15), exret(32), logsale(17), price(31), logat(16), cashmta(4), rsize(30), cashat(3), sigma(33), chat(5), chlct(6), actlct(1), saleat(27), lctsale(13), qalct(24), ltat(14), fat(7), apsale(2), lctat(11), invtsale(9).}
    \label{fig_shap}
\end{figure}

\begin{figure}[t!]
    \centering
    \begin{subfigure}{0.48\textwidth}
        \includegraphics[scale=0.37]{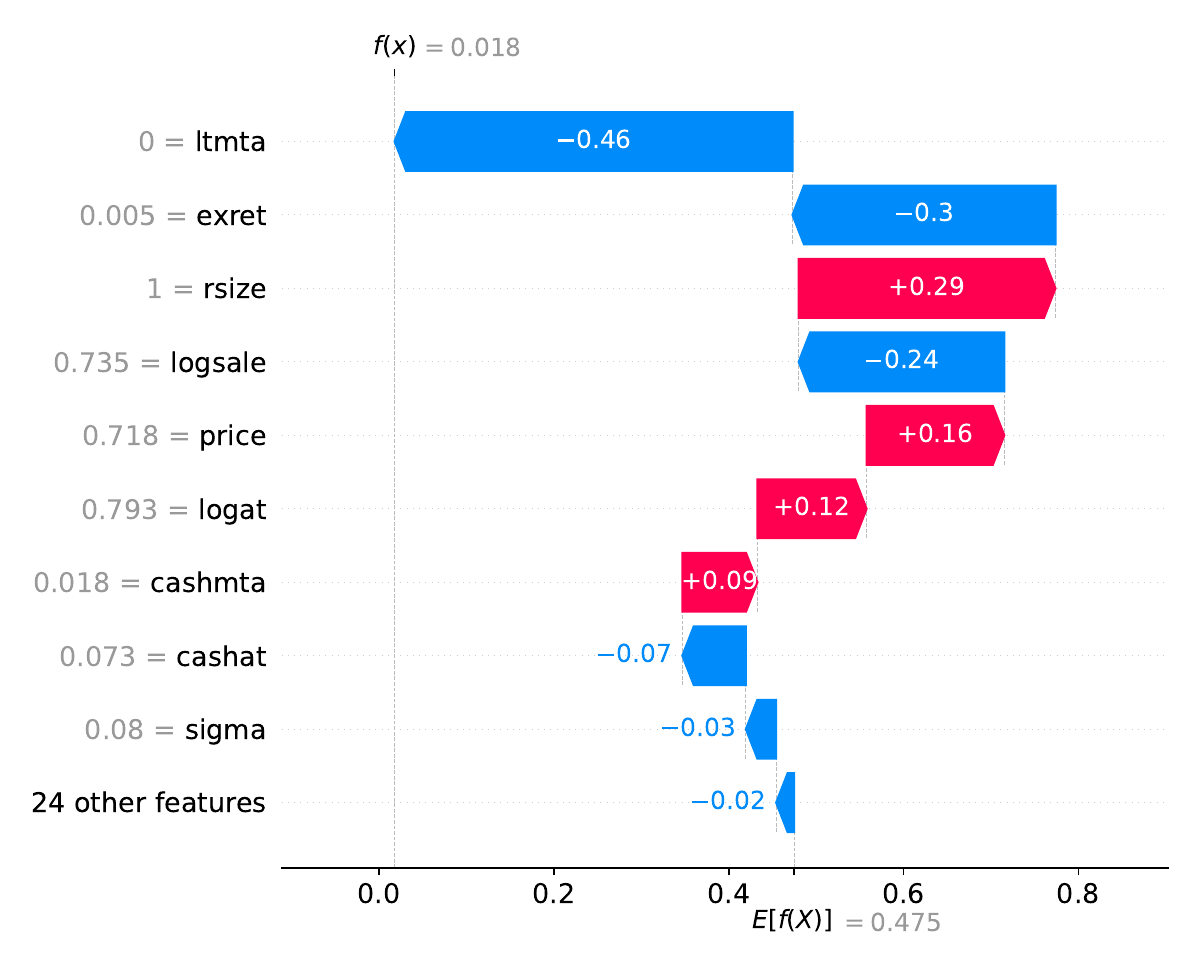}
    \end{subfigure}
    \begin{subfigure}{0.48\textwidth}
        \includegraphics[scale=0.37]{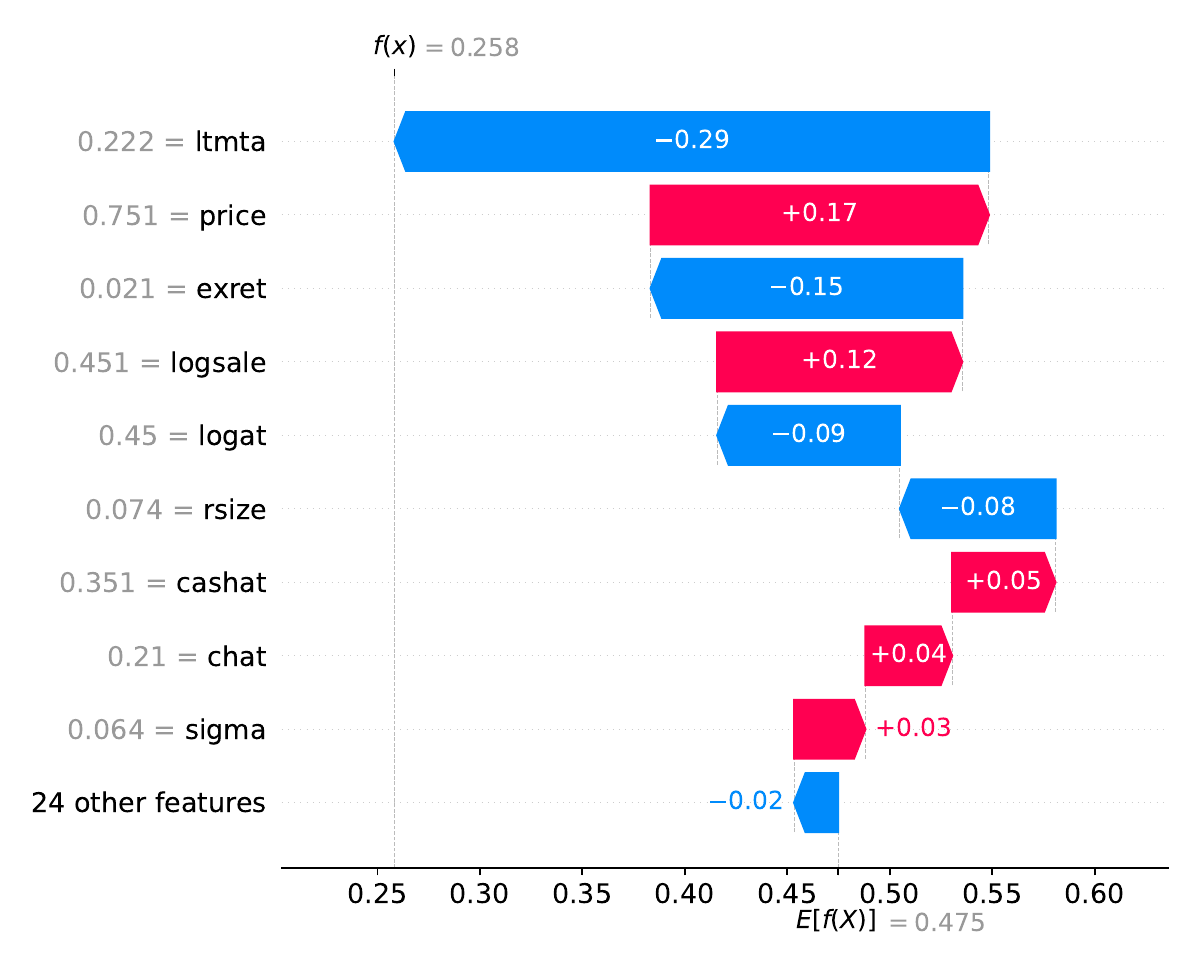}
    \end{subfigure}
    \caption{Effect of different variables on the probability of bankruptcy. Each arrow shows the effect on the probability of bankruptcy, starting from a common baseline. The variable names in Table \ref{tbl_variables} corresponding to the above variable codes are given in parentheses as follows: lmta(15), exret(32), logsale(17), price(31), logat(16), cashmta(4), rsize(30), cashat(3), sigma(33), chat(5).}
    \label{fig_shap2}
\end{figure}
We estimate the SHAP values for the test set using the kernel SHAP method introduced in \cite{lundberg2017unified} and utilize the python library developed by the same authors\footnote{\href{https://github.com/slundberg/shap}{https://github.com/slundberg/shap}}. Figure \ref{fig_shap} shows SHAP values for the 20 variables with the greatest impact on 1-year predictions (left panel). Each point denotes an SHAP value for one specific prediction. The x-axis shows the SHAP values and the color represents the values of the feature from low to high. As can be seen in the figure, the variable ltmta = total liabilities/(market equity + total liabilities) has the greatest impact on the classification performance. The three panels on the right show the relationship between the three main variables with the greatest impact on 1-year bankruptcy predictions and their corresponding SHAP value. For both ltmta and excess return, the SHAP values increase as the variables increase, while for logsale = log(Sales) the SHAP values decreases as logsale increases. \review{It should be noted that the SHAP values presented in this section should be taken with some caution, as some of the $\xo$ variables are correlated (see Appendix \ref{app_corr}), while the SHAP values are computed under the assumption of uncorrelated predictor variables. Hence, the ranking of the variables with the greatest influence on the predictions might be misleading.}

Finally, Figure \ref{fig_shap2} shows the effect of the 9 most important variables in the modality $\xo$ that drives the probability of bankruptcy for 2 different companies in the test set. Each arrow shows the effect on the probability of bankruptcy, starting from a common baseline. 

\subsection{Generative Model}\label{sec_generative}
The CMMD model is not only a discriminative model, but also a generative model capable of doing cross-modal generation. That is, using the available modalities $\xo$ at the test time, the CMMD generates joint latent representations from the prior distribution $p(\z|\xo)$, which are later used in the generative model of the missing modality $p(\xm|\xo,\z)$ (see Figure \ref{fig_cmmd}). In other words, the CMMD model generates the missing MDA modality conditioned on the available modalities, accounting and market data, and the joint representation of all modalities $\z$. In this particular case, the CMMD model generates TF-IDF scores, which represent the tokens that are most important in a given MDA text, since we converted MDAs to TF-IDF scores for model training. 

\begin{table}[t!]
\def\arraystretch{1.1}
\setlength{\tabcolsep}{3.5pt}
\scriptsize
    \centering
    \caption{Tokens corresponding to the highest TF-IDF scores generated with the model $p(\xm|\xo,\z)$ that are part of the \cite{loughran2011liability} financial tone dictionary. The number of companies that include a particular token is shown in parenthesis.}
    \begin{tabular}{lllll}
        \multicolumn{5}{c}{\textbf{Negative words present in the 500 companies with the highest probability of bankruptcy}}  \\ \cmidrule(lr){1-5} 
        %Word (counts) & Word (counts) & Word (counts) & Word (counts) & Word (counts) \\
        accident (500)&erroneous (500)&relinquish (500)&suffered (243)&expose (12)\\
        accidental (500)&erroneously (500)&relinquished (500)&lie (162)&exposed (12)\\
        argue (500)&error (500)&relinquishment (500)&infringes (142)&impracticable (9)\\
        argued (500)&ill (500)&retaliation (500)&refuse (133)&disadvantage (8)\\
        bankrupt (500)&impede (500)&stolen (500)&refused (133)&erring (6)\\
        bankruptcy (500)&impeded (500)&strain (500)&lose (67)&inadequacy (4)\\
        barred (500)&impediment (500)&strained (500)&erred (64)&forfeit (3)\\
        circumvent (500)&inappropriate (500)&subjecting (500)&pretrial (62)&forfeiture (3)\\
        circumvented (500)&inappropriately (500)&sue (500)&persistent (60)&monopoly (3)\\
        circumvention (500)&incident (500)&sued (500)&cease (56)&resign (2)\\
        contend (500)&incorrect (500)&threat (500)&ceased (56)&resignation (2)\\
        corrupt (500)&infringe (500)&insolvent (499)&damaging (40)&complain (2)\\
        cut (500)&interrupt (500)&incomplete (493)&overestimate (33)&encumbrance (2)\\
        decline (500)&interrupted (500)&preclude (492)&reject (31)&force (2)\\
        declined (500)&lag (500)&turmoil (491)&rejected (31)&layoff (2)\\
        depletion (500)&lagged (500)&incidence (490)&contends (30)&standstill (2)\\
        destroy (500)&lagging (500)&intermittent (481)&underfunded (22)& \\
        destroyed (500)&late (500)&impossible (478)&lying (21)& \\
        disagree (500)&malfunction (500)&exploit (473)&terminate (21)& \\
        disagreement (500)&miss (500)&persist (470)&terminates (21)& \\
        disciplinary (500)&overstate (500)&quit (467)&erode (20)& \\
        dishonest (500)&posing (500)&misrepresent (268)&eroded (20)& \\
        dishonesty (500)&prevents (500)&misrepresented (268)&corrected (19)& \\
        dissatisfaction (500)&recession (500)&liquidation (257)&damage (19)& \\
        enjoin (500)&recessionary (500)&suffer (243)&complication (14)& \\
        & & & & \\
        \multicolumn{5}{c}{\textbf{Positive words present in the 500 companies with the lowest probability of bankruptcy}}  \\ \cmidrule(lr){1-5}
        able (500)&improve (500)&satisfaction (500)&regained (443)&collaborating (9)\\
        abundant (500)&improved (500)&exclusivity (497)&friendly (341)&advantage (8)\\
        accomplish (500)&improvement (500)&improving (497)&win (92)&assure (5)\\
        accomplishment (500)&improves (500)&gain (493)&despite (34)&strength (4)\\
        collaborative (500)&outperform (500)&gains (447)&perfect (16)&strengthen (4)\\
        confident (500)&rebound (500)&gained (443)&perfected (16)&solving (2)\\
        exciting (500)&rebounded (500)&regain (443)&good (14)& \\
    \end{tabular}
    \label{tbl_words_dict}
\end{table}
% dictionary last updated with neg words 2014
We generate the $\xm$ modality for all 40,950 companies in the test period from 1994 to 2020. We then rank all companies according to their probability of bankruptcy estimated with the trained CMMD model. We select 500 companies with the highest and lowest probability of bankruptcy and obtain the 1,000 tokens that are the most important according to the TF-IDF score generated by the generative model $p(\xm|\xo,\z)$. Finally, we check which of the 1 million generated tokens, corresponding to all companies that we selected, are present in the negative and positive tone dictionaries introduced in \cite{loughran2011liability}. \cite{loughran2011liability} analyze the tone of financial words in 50,115 Form 10-K documents between 1994 and 2008, and they create six different dictionaries of words that have negative, positive, uncertain, litigious, strong-modal, and weak-modal implications. Their empirical results show that companies with fillings including a high percentage of negative words, on average, have negative returns on the days around the 10-K filling date. 

\begin{figure}[t]
    \centering
    \begin{subfigure}{0.48\textwidth}
    \centering
        \includegraphics[scale=0.45]{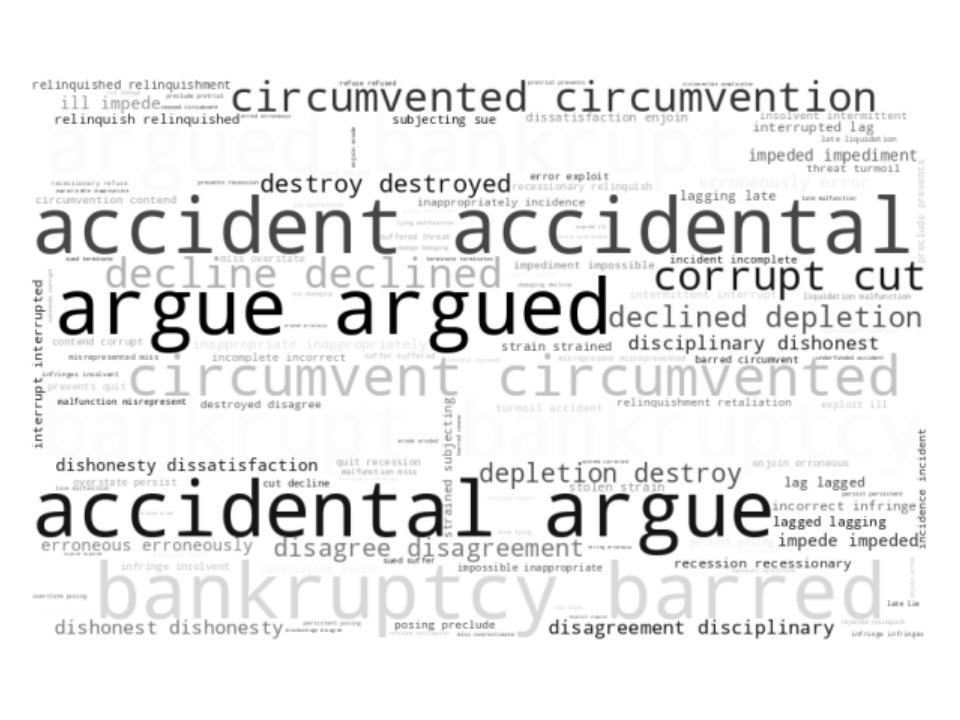}
        \caption{Companies with highest probability of bankruptcy}
    \end{subfigure}
    \begin{subfigure}{0.48\textwidth}
    \centering
        \includegraphics[scale=0.45]{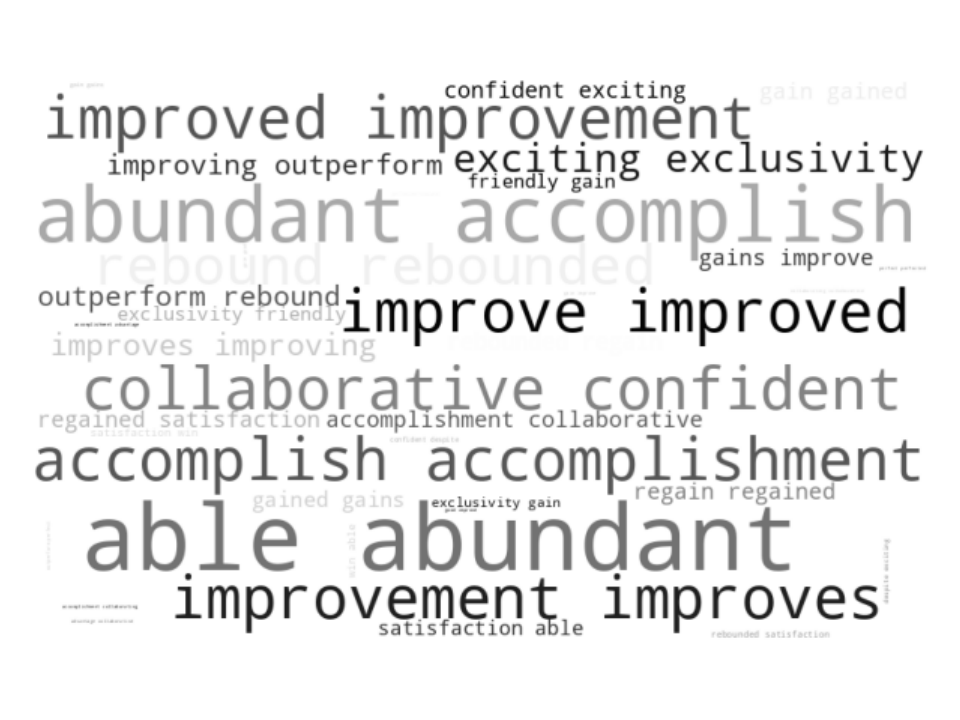}    
        \caption{Companies with lowest probability of bankruptcy}
    \end{subfigure}
    \caption{
    Wordcloud of tokens corresponding to TF-IDF scores generated with the generative model $p(\xm|\xo,\z)$. 
    }
    \label{fig_wordcloud}
\end{figure}

%corresponding to the 1,000 highest TF-IDF scores generated
Table \ref{tbl_words_dict} shows which of the 1 million tokens we generated with the model $p(\xm|\xo,\z)$ are also part of the negative (top panel) and positive (bottom panel) tone dictionaries in \cite{loughran2011liability}. For example, the word \textit{accident}, which is part of the negative tone dictionary, is among the 1,000 highest TF-IDF scores in all 500 selected companies with the highest probability of bankruptcy. On the other hand, the word abundant is among the 1,000 highest TF-IDF scores for the 500 companies with the lowest probability of bankruptcy, which happens to be also in the positive tone dictionary. Figure \ref{fig_wordcloud} shows wordcloud images for all words and counts in Table \ref{tbl_words_dict}.

\subsection{Latent Representations of Companies}\label{sec_representations}
In order to better understand the multimodal representations $\z$, it is possible to visualize both the latent variables of the prior distribution $\z \sim p(\z|\xo)$ and the time evolution of its parameters $\bm{\mu}$ and $\bm{\Sigma}$, which is a diagonal matrix with parameters $\bm{\sigma}$.

To visualize latent variables $\z$, we use the CMMD model from Experiment I to generate latent representations from the prior distribution $p(\z|\xo)$ for the period 2017 to 2020. In this case, the dimension of the latent space is 50 (see Appendix \ref{app_grid}), i.e. $\z \in \mathbb{R}^{50}$, hence we use t-SNE \citep{van2008visualizing} to obtain two-dimensional vectors that can be visualized. 

\begin{figure}[!t]
    \centering
    \begin{subfigure}{0.32\textwidth}
    \includegraphics[scale=0.37]{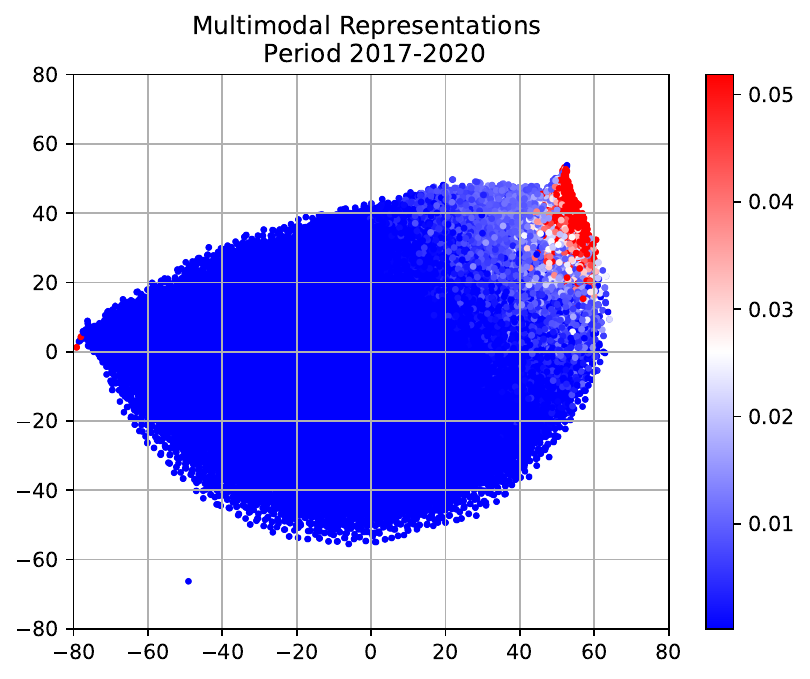}
    \caption{}
    \end{subfigure}
    \begin{subfigure}{0.32\textwidth}
    \includegraphics[scale=0.37]{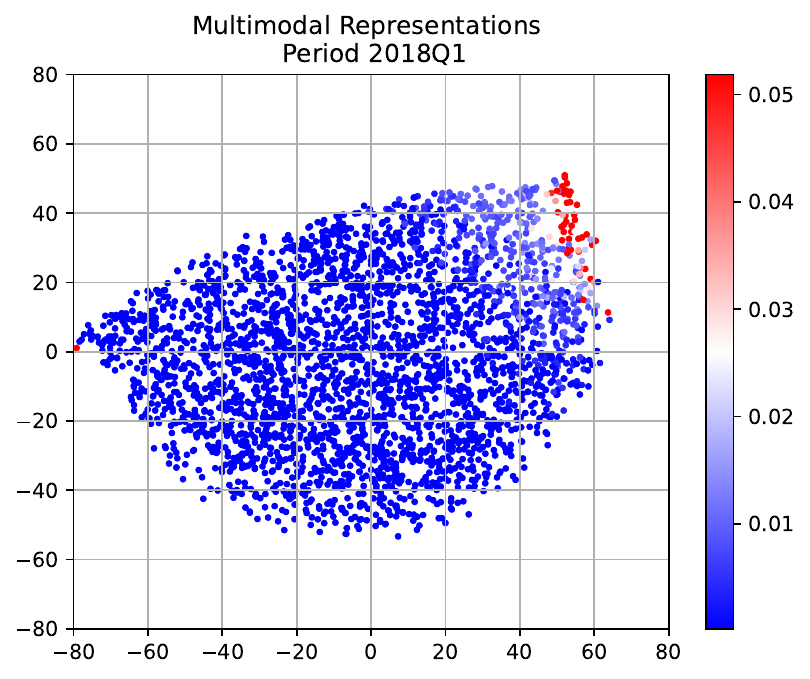}
    \caption{}
    \end{subfigure}
    \begin{subfigure}{0.32\textwidth}
    \includegraphics[scale=0.37]{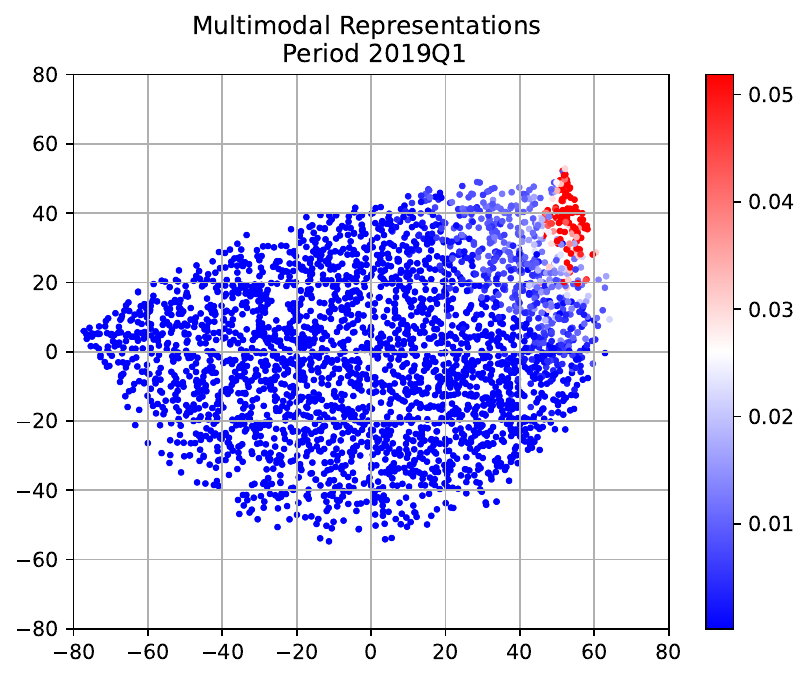}
    \caption{}
    \end{subfigure}
    \caption{Latent representations for (a) the period 2017-2020, (b) the first quarter 2018, and (c) the first quarter 2019. The color of the scatter corresponds to the bankruptcy probability estimated by the CMMD model.}
    \label{fig_datareps}
\end{figure}

Panel (a) in Figure \ref{fig_datareps} shows data representations for all companies during the entire test period, where the scatter color is given by the probability of bankruptcy estimated by the CMMD classifier. It is interesting to see that companies with relatively high probability of bankruptcy cluster in the upper-right corner. Further, panels (b) and (c) show representations during the first quarters of 2018 and 2019, respectively. Note that there are relatively fewer companies where the CMMD model estimates a high probability of bankruptcy during 2018Q1. From all figures, it is interesting to see that the estimated bankruptcy probability shows a smooth transition across the two-dimensional space. This suggests that the latent representations of the CMMD model are capable of capturing the spatial coherence \citep{bengio2013representation} of financial distress, which means that spatially near-by observations tend to be associated with the same value of probability of bankruptcy.

To follow the development over time of the parameters $\bm{\mu}$ and $\bm{\sigma}$, we aggregate quarterly latent variables $\z$ for all firms as follows: Let $\z^k \sim \mathcal{N}(\bm{\mu}^k,\bm{\Sigma}^k)$ be the latent representation for the the \textit{k}th company. Given that all covariance terms are 0, the standard deviation (std) of the sum of all variables in $\z^k$ is $std(z_1^k+z_2^k+\cdots+z_d^k) = \sqrt{\sum_{i=1}^d var(z_i^k)}$, where $d$ is the dimension of $\z$. Therefore, we define an aggregated $\bm{\sigma}^*$ parameter for all companies as
\begin{equation}\label{eqsigma}
    \bm{\sigma}^* = \sum_{k=1}^K \sqrt{\sum_{i=1}^d var(z_i^k)},
\end{equation}
where $K$ is the total number of companies. Following the same logic, we define an aggregated $\bm{\mu}^*$ parameter for all companies as
\begin{equation}\label{eqmu}
    \bm{\mu}^* = \sum_{k=1}^K \sum_{i=1}^d \mu_i^k.
\end{equation}
For each quarter $q$ and company \textit{k}, we generate multimodal representations $\z_k^q \sim p(\z|\xo)$ and calculate yearly representations $\z_k^Y= \dfrac{1}{Q}\sum_{q=1}^{Q}\z_k^q$, where $Y$ denotes a given year and $Q$ is the latest observed quarter in $Y$. Finally, we calculate the aggregate yearly parameters $\bm{\sigma}_Y^*$ and $\bm{\mu}_Y^*$ using Equations \ref{eqsigma} and \ref{eqmu}. 

\begin{figure}[t!]
    \centering
    %\begin{subfigure}{0.48\textwidth}
    %    \includegraphics[scale=0.5]{evolutio.pdf}        
    %    \caption{}
    %\end{subfigure}
    %\begin{subfigure}{0.48\textwidth}
        \includegraphics[scale=0.4]{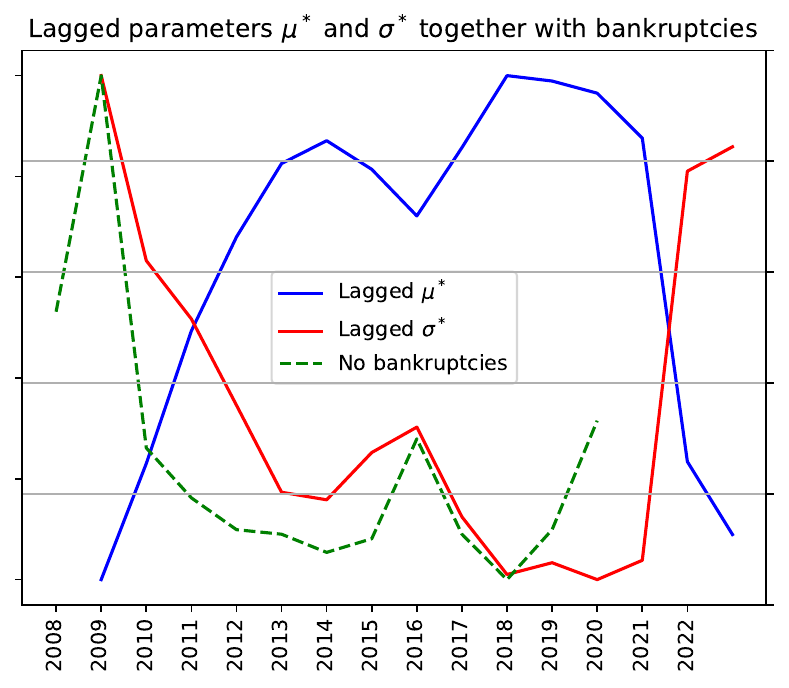}        
    %    \caption{}
    %\end{subfigure}
    %\caption{Time evolution of the aggregated parameters $\bm{\mu}^*$ and $\bm{\sigma}^*$ (a) and lagged parameters $\bm{\mu}^*$ and $\bm{\sigma}^*$ together with number of bankruptcies (b).}
    \caption{Time evolution of lagged parameters $\bm{\mu}^*$ and $\bm{\sigma}^*$ together with number of bankruptcies.}
    \label{fig_evol}
\end{figure}

%Panel (a) in Figure \ref{fig_evol} shows the evolution of these parameters. It is interesting to see that as $\bm{\mu}_Y^*$ increases, its uncertainty $\bm{\sigma}_Y^*$ decreases. Therefore, on average, the prior distribution for the \textit{k}th company has smaller standard deviation for higher values of the mean.
We calculate $\bm{\mu}_Y^*$ and $\bm{\sigma}_Y^*$ for all years between 2008 to 2022 using a CMMD model trained with data from 1994 to 2007.  Figure \ref{fig_evol} compares the lagged parameters $\bm{\mu}^*$ and $\bm{\sigma}^*$ with the number of bankruptcies from 2008 to 2020. This comparison corresponds to the scenario of 1-year bankruptcy prediction, in which latent representations from the prior distribution $\z \sim p(\z|\xo)$ during 2008 are used to forecast bankruptcies in 2009, for example. We can see that the aggregated parameters $\bm{\mu}^*$ and $\bm{\sigma}^*$ are negatively and positively correlated with the number of bankruptcies, respectively. The correlation factor between $\bm{\mu}^*$ and the number of bankruptcies is -0.77 and between $\bm{\sigma}^*$ and the number of bankruptcies is 0.78. This correlation explains why multimodal representations $\z$ stand out as the best predictor of corporate bankruptcy, as shown in the experiments of this research.

\section{Conclusion}\label{sec_conclusion}
\review{We use the CMMD model to solve the limitation of traditional corporate bankruptcy models using MDA data. Traditional models would only make predictions for 60\% of the firms, as the remaining 40\% of the companies lack the MDA section according to our data sample. Corporate bankruptcy prediction with the CMMD model is based on multimodal learning, which is a method that learns joint latent representations (from accounting, market, and textual data) that can be used to predict bankruptcy and generate the missing MDA modality in this case. To our knowledge, this research introduces for the first time the concept of multimodal learning in bankruptcy prediction models.}

The empirical results of this research show that if the training data set is large enough, our proposed methodology outperforms a large number of traditional classifier models. \review{Furthermore, we show that the CMMD model is able to generate words from the missing MDA data that actually belong to the financial tone dictionaries used in previous literature. Finally, using SHAP values, we show that the bankruptcy predictions of our proposed method can be interpreted.}

\section{Aknowledgments}
We thank Sudheer Chava and Feng Mai for sharing the bankruptcy and MDA data with us, respectively.

\section*{Appendices}
\renewcommand{\thesubsection}{\Alph{subsection}}
\renewcommand{\thetable}{A\arabic{table}}
\setcounter{table}{0}

\subsection{Hyperparameters Optimization}\label{app_grid}
We fine-tune the hyperparameters of all models used in this research. Table \ref{tbl_gridsearch} shows the parameters considered in the grid search approach. 

\begin{table}[t!]
\footnotesize
\centering
\caption{Grid search for hyperparameter tuning for all models studied in this research. We consider all combinations of the hyperparameters listed here. The superscripts *, **, and *** denote the best architecture for 1, 2, and 3 years bankruptcy predictions, respectively (see Table \ref{tbl_aucresults}). Likewise, the subscript **** denotes the best architecture used in the experiments shown in Table \ref{tbl_aucresults2}. For the rest of hyperparamters in the benchmark models, we used default values in the \texttt{sklearn} implementation.}
\def\arraystretch{1}
\setlength{\tabcolsep}{15pt}
\begin{tabular}{|c|c|}
\hline
\textbf{Hyperparameter}     & \textbf{CMMD}          \\ 
\hline
latent dimension  & 50$^{*,**,***}$, 100, 150, 250$^{****}$ \\
dropout encoder, prior, decoder       & 0$^{***,****}$, 0.1$^{*,**}$ \\
layer size classifier & [50,50]$^{***}$, [100,100]$^{****}$, [150,150]$^{*,**}$ \\
layer size encoder, prior, decoder & [50,50,50]$^{****}$, [100,100,100]$^{*,**,***}$ \\
omega   & 0.25, 0.5, 0.75$^{*,**,***}$, 0.9$^{****}$ \\
\hline
&   \textbf{RF} \\
\hline
no. of trees & 50$^{*,**}$, 100$^{***}$, 150, ..., 450$^{****}$, ..., 900, 950, 1000\\
max. depth of the tree & 10$^{****}$, 20$^{**}$, ..., 40$^{*}$, 50$^{***}$, ..., 100, 110 \\
no. of features when splitting & auto$^{*,**,****}$, sqrt$^{***}$ \\
min. no. of samples to split an internal node & 2, 5$^{**,***,****}$, 10$^{*}$ \\
min. no. of samples to be at a leaf node & 1$^{**,***}$, 2, 4$^{*,****}$ \\
bootstrap & True$^{*,**,***,****}$, False \\
\hline
 & \textbf{SVM} \\
 \hline
 regularization parameter & 0.1, 0.5, 1, 5$^{****}$, 10$^{*,**,***}$ \\
 kernel & linear, rbf$^{*,**,***,****}$ \\
 kernel coefficient & scale$^{*,**,***,****}$, auto \\
 \hline
 & \textbf{MLP} \\
 \hline
 layer size & 10$^{****}$, 20, 50, 100, [10,10], [20,20]$^{*,***}$, [50,50], [100,100]$^{**}$\\
 activation function & logistic, tanh$^{***}$, relu$^{*,**,****}$ \\
 learning rate & constant$^{*,****}$, invscaling$^{***}$, adaptive$^{**}$ \\
 L2 regularization & 0.0001$^{*,**}$, 0.001$^{***}$, 0.01, 1$^{****}$ \\
 \hline
 & \textbf{k-NN} \\
 \hline 
 no. of neighbors & 3,5,10,15,20,30$^{***}$,50,80,110$^{****}$,150$^{*,**}$,200,300 \\
 weight function for predictions & uniform, distance$^{*,**,***,****}$ \\
 distance metric & euclidean$^{****}$, manhattan$^{*,**,***}$, minkowski \\
 \hline
 & \textbf{NB} \\
 \hline
 prior class probabilities & [0.5, 0.5]$^{*,***,****}$, [0.97, 0.03]$^{**}$ \\
 \hline
 & \textbf{LR} \\
 \hline
 penalty  & l2, l1$^{*,**,***,****}$ \\
 inverse of regularization & 0.0001, 0.001, 0.01, 0.1, 0.5$^{****}$, 1, 5, 10$^{*}$, 15, 20, 50$^{**,***}$ \\
 \hline
\end{tabular}
\label{tbl_gridsearch}
\end{table}

\subsection{Correlation Matrix}\label{app_corr}
\renewcommand{\thefigure}{B\arabic{figure}}
\setcounter{figure}{0}
Figure \ref{fig_correlation} shows correlation values for all variables in the $\xo$ modality.  The variable names in Table \ref{tbl_variables} corresponding to the variable codes in the correlation diagram are given in parentheses as follows: 
actlct (1), 
apsale (2),
cashat (3), 
cashmta (4),
chat (5), 
chlct (6),
fat (7), 
invchy/saley (8),
invtsale (9),
lctchat (10),
lctat (11), 
lctlt (12),
lctsale (13),
ltat (14),
lmta (15),
logat (16),  
logsale (17),
mb (18),
niat (19),
nimta (20),
nisale (21),
oiadpat (22),
oiadpsale (23),
qalct (24),   
reat (25),
relct (26),
saleat(27),  
seqat (28),
wcapat (29),
rsize (30),
price (31),
exret (32),   
sigma (33). 

\begin{figure}[t!]
    \centering
    \includegraphics[scale=0.37]{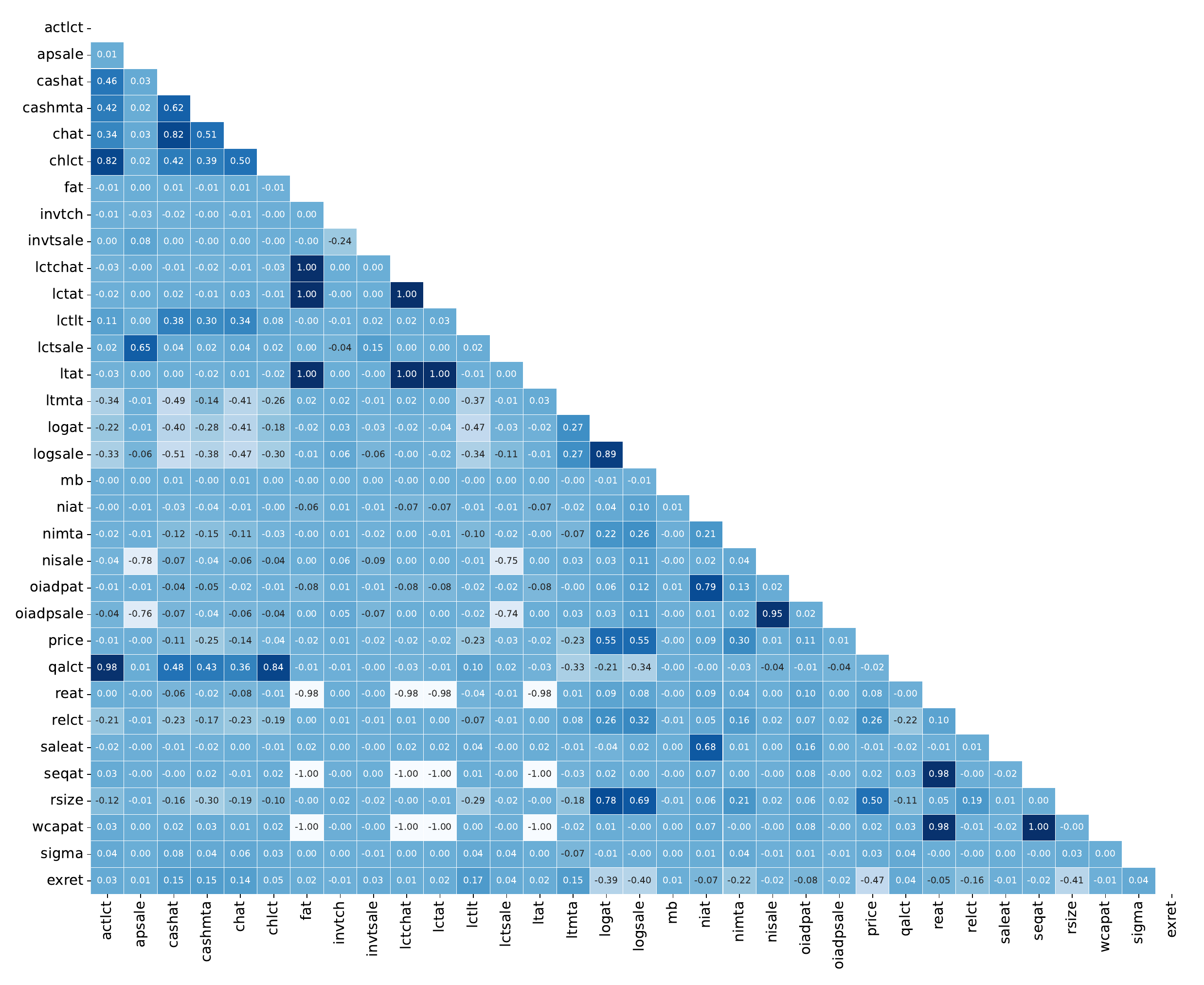}
    \caption{Correlation values for all variables in the $\xo$ modality.}
    \label{fig_correlation}
\end{figure}

\clearpage
\bibliographystyle{myapalike}
\bibliography{bibliography}

\end{document}